\documentclass{appolbHEP}  %%FRK: \runhead changed, for preprint, out for APPB

\usepackage{epsfig,amsmath,amssymb,amsfonts,xspace}   %%FRK

\newcommand {\version}{v5}
\newcommand{\bdi}{\begin{displaymath}}
\newcommand{\edi}{\end{displaymath}}
\newcommand{\bfi}{\begin{figure}}
\newcommand{\efi}{\end{figure}}

\newcommand{\beq}{\begin{equation}}
\newcommand{\eeq}{\end{equation}}
\newcommand{\beqa}{\begin{eqnarray}}
\newcommand{\eeqa}{\end{eqnarray}}
\newcommand{\no}{\nonumber}

\newcommand{\DS}[1]{$\mathsf{#1}$\xspace}       % discrete symmetry, as in Weinberg's book
\newcommand{\action}{\mathcal{S}}               % action S
\newcommand{\plus}{\oplus}                      % plus photon mode
\newcommand{\minus}{\ominus}                    % minus photon mode
\newcommand{\msmall}{m}                         % CS-mass scale

\newcommand{\ii}{\mathrm{i}}                    % imaginary i
\newcommand{\BigO}{\mathrm{O}}                  % order O
\newcommand{\fracnew}[2]{\frac{{#1}_{\vphantom{!_a}}}{{#2}^{\vphantom{a}}}}
\def\openone{\leavevmode\hbox{\small1\kern-3.8pt\normalsize1}} % from revtex3

\newcommand{\ve}{\mathbf{e}}
\newcommand{\va}{\mathbf{a}}
\newcommand{\vb}{\mathbf{b}}
\newcommand{\vx}{\mathbf{x}}
\newcommand{\vy}{\mathbf{y}}
\newcommand{\vk}{\mathbf{k}}
\newcommand{\vm}{\mathbf{m}}
\newcommand{\vn}{\mathbf{n}}
\newcommand{\vp}{\mathbf{p}}
\newcommand{\vq}{\mathbf{q}}
\newcommand{\vv}{\mathbf{v}}
\newcommand{\vE}{\mathbf{E}}
\newcommand{\vB}{\mathbf{B}}
\newcommand{\vxi}{\boldsymbol{\xi}} %amsmath

\newcommand{\qp}{q_\parallel}
\newcommand{\qpabs}{|q_\parallel|}

\newcommand{\rmd}{\mathrm{d}}
\newcommand{\dx}{\!\mathrm{d}^4x\,\,}
\newcommand{\dint}[1]{\!\mathrm{d}#1\,\,}

\newcommand {\LC}     {Levi--Civita}
\newcommand {\CS}     {Chern--Si\-mons}
\newcommand {\MCS}    {Max\-well--Chern--Si\-mons}
\newcommand {\bcs}    {boundary conditions}

\newcommand {\qfth}   {quantum field theory}

\newcommand {\YM}     {Yang--Mills}

\newcommand {\cgth}   {chiral gauge theory}

\newcommand {\clgth}  {chiral lattice gauge theory}

\newcommand {\rhs}    {right-hand side}
\newcommand {\SM}     {Standard Model}

\newcommand{\lgamma}  {l_\gamma}

\begin{document}
\eqsec

\title{\vspace*{-1.25cm}%%FRK: for preprint, out for APPB
       NONTRIVIAL SPACETIME TOPOLOGY,\vspace*{1mm}\\
       CPT VIOLATION, AND PHOTONS\thanks{Published in:
       \emph{CP Violation and the Flavour Puzzle:
       Symposium in Honour of Gustavo C. Branco},
       edited by D. Emmanuel-Costa,
       L. Lavoura, F. Mota, P.A. Parada, M.N. Rebelo, J.I. Silva-Marcos,
       Krak\'{o}w, Poligrafia Inspektoratu, 2005, pp. 157--191;
       hep-ph/0511030 (\version)}%%KA--TP--12--2005;
       }

\author{Frans R. Klinkhamer
\address{Institute for Theoretical Physics, University of Karlsruhe (TH)\\
         76128 Karlsruhe, Germany}
}
\maketitle
\begin{abstract}
A physical mechanism for \DS{CPT} violation is reviewed, which relies on
chiral fermions, gauge interactions, and nontrivial spacetime topology.
The nontrivial topology can occur at the very largest scale
(\eg, at the ``edge'' of the universe) or at the very smallest scale
(\eg, from a hypothetical spacetime foam).
The anomalous effective gauge field action includes, most likely, a
\DS{CPT}--odd Chern--Simons-like term. Two phenomenological
photon models with Abelian  Chern--Simons-like terms are discussed.
\end{abstract}

\section{Introduction}
\label{Introduction}

The \DS{CPT} ``theorem'' \cite{L54,P55,B55,L57,J57}
states that any local relativistic \qfth~is invariant under the combined
operation of charge conjugation (\DS{C}),
parity reflection (\DS{P}), and time reversal (\DS{T}),
in whichever order. Considered by itself,
the theorem  is based on the following main assumptions (cf. Ref.~\cite{L57}):
\begin{itemize}
\vspace*{-0.00\baselineskip}
\item Minkowski spacetime, with manifold $\mathbb{R}^4$ and flat
      metric $\eta_{\mu\nu}$;\vspace*{-0.25\baselineskip}
\item invariance under transformations of the proper
      or\-tho\-chro\-nous Lo\-rentz group ${\mathcal L}_+^\uparrow$ and
      spacetime translations;\vspace*{-0.25\baselineskip}
\item normal spin--statistics connection;\vspace*{-0.25\baselineskip}
\item locality and Hermiticity of the Hamiltonian.
\vspace*{-0.00\baselineskip}
\end{itemize}
A detailed discussion of the theorem can be found in, \eg,
Refs. \cite{SW64,J65,BLOT90}
and some of its consequences have been reviewed in, \eg,
Refs. \cite{S64,TD86,BLS99}.

Here, we go further and ask the following question:
\emph{\underline{can} \DS{CPT} invariance be violated at all
in a physical theory and, if so, \underline{is} it in the real world?}
It is obvious that something  ``out of the ordinary''
is required for this to be the case.
Two possibilities, in particular, have been discussed in the literature.

First, there is \emph{quantum-gravity theory},
which may or may not lead to \DS{CPT} violation; cf. Refs.~\cite{W80,H85}.
The point is, of course, that Lorentz invariance does not
hold in general. Still, a \DS{CPT} theorem can be ``proven,''
in the Euclidean  formulation, for asymptotically-flat spacetimes \cite{AG83}.
In the canonical formulation, on the other hand, certain
semiclassical (weave) states could affect the Lorentz invariance of
Maxwell theo\-ry at the Planck scale and break  \DS{CPT}
invariance \cite{GP99,AMU02}.
But, at the moment, this is not a firm prediction, especially as the
complete theory is not formulated  \cite{CorichiSudarsky,Nicolai}.

Second, there is \emph{superstring theory},
which may or may not give \DS{CPT} violation;
cf. Refs.~\cite{S89,KP91,Polchinski}.
The point, now, is the (mild) nonlocality of the theory.
There exists, however, no convincing calculation showing the
necessary violation of \DS{CPT}.
And, here also, the complete theory is not formulated
\cite{GreensiteKlinkhamer,Mandelstam,Rovelli}.

In this contribution, we discuss a third possibility:
\emph{certain spacetime topologies and classes of
chiral gauge theories have Lorentz and \DS{CPT} invariance
necessarily broken by quantum effects.}
The main article on this ``\DS{CPT} anomaly'' is Ref. \cite{K00},
which, under certain assumptions,
finds a \DS{CPT}--odd Chern--Simons-like term in the
effective gauge field action.
(The connection with earlier work on sphalerons, spectral flow, and anomalies
is explained in Refs.~\cite{K98,KlinkhamerRuppJMP2003}.)
Further aspects of the \DS{CPT} anomaly have been discussed in
Refs. \cite{KN01,KM01,KlinkhamerSchimmel,KlinkhamerRuppPRD2004}.
The corresponding Maxwell--Chern--Simons model
(standard electrodynamics with an Abelian Chern--Simons-like term
added to the action) has been studied
in Refs.~\cite{AK01npb,AK01plb,AK03,K02,KaufholdK05,KantK05}
and a related model with random coupling constants in
Refs.~\cite{KlinkhamerRuppPRD2005,KlinkhamerRuppNewAstRev2005}.
Here, we intend to summarize the main results and to
point out some of the important open questions.

The outline of the present article is as follows.
In Sec.~\ref{Example}, a  realistic example of
a theory with anomalous \DS{CPT} violation
is given, together with a heuristic argument for the origin of the effect.

In Sec.~\ref{Exact2D},  the \DS{CPT} anomaly is established for a class of
exactly solvable two-dimensional theories (the details are relegated to
Appendix~\ref{Effective action 2D}).
In Sec.~\ref{Nonperturbative4D},  the existence of a \DS{CPT} anomaly is
shown nonperturbatively for a particular formulation of
four-dimensional chiral lattice gauge theory
(the main steps are sketched in Appendix~\ref{CPT anomaly on a 4D  lattice}).
In Sec.~\ref{Perturbative4D},
the \DS{CPT} anomaly is obtained perturbatively for a class of
four-dimensional chiral gauge theories, which includes the example of
Sec.~\ref{Example}. Two types of space manifolds are considered explicitly,
a cylindrical manifold with nontrivial topology at the largest scales
and a ``punctured'' manifold with
nontrivial topology at the smallest scales.

In Sec.~\ref{MCSmodel+phenomenology},
the phenomenological Maxwell--Chern--Simons model
(corresponding to the anomalous effects of a cylindrical manifold)
is reviewed, while the important
issue of microcausality is dealt with in Appendix~\ref{Microcausality in 4D MCS}.
The model Chern--Simons-like term  modifies the propagation of photons,
which may be relevant to photons traveling over cosmological
distances (and, possibly, to the origin of the big bang).
With suitable interactions added, further
effects appear such as vacuum Che\-ren\-kov radiation and
photon triple-splitting. In curved spacetime backgrounds,
other novel phenomena occur such as stable orbits of
light around a nonrotating central mass and gravitational-redshift
splitting between the two polarization modes.

In Sec.~\ref{Random-coupling model},
the phenomenology of a random-coupling photonic model
(corresponding to the anomalous effects of a punctured manifold)
is discussed. The resulting dispersion law has been calculated in the
long-wavelength limit and  can be confronted with
high-energy-astrophysics data to
constrain (or determine) the parameters of the photon model considered.

In Sec.~\ref{Conclusion},  some concluding remarks are presented.

For the benefit of the reader, we note that this
review article essentially consists of two tracks, apart from
Secs.~\ref{Example} and \ref{Conclusion} with general comments.
The first track focuses on the basic physics of the \DS{CPT} anomaly
and consists of Secs.~\ref{Exact2D}, \ref{Nonperturbative4D},
and \ref{Perturbative4D}.
The second track discusses nonstandard photon physics from two simple
phenomenological models (the Maxwell--Chern--Simons model proper
and a related photon model with random coupling constants)
and consists of Secs.~\ref{MCSmodel+phenomenology}
and \ref{Random-coupling model}.
Both tracks are more or less independent, but the second one is, of course,
motivated by the first.

\section{Example and heuristics}
\label{Example}

The anomalous \DS{CPT} violation mentioned in the Introduction
is perhaps best illustrated by  a concrete example.
Consider the following four-dimensional
spacetime manifold $\mathsf{M}^\prime$ with metric $g_{\mu\nu} (x)$
and vierbeins $e_\mu^a (x)$:
\beq \label{R3S1}
\big(\,\mathsf{M}^\prime \,;\, g_{\mu\nu} (x) \,;\, e_\mu^a (x) \,\big) =
\big(\,
{\mathbb R}^3 \times S^1_\text{\,PSS} \,;
\,\eta_{\mu\nu} \,;
\,\delta_\mu^a\,
\big)\;,
\eeq
for Minkowski tensor $(\eta_{\mu\nu})=\mathrm{diag}(1,-1,-1,-1)$,
Kronecker symbol $\delta_\mu^a$, and coordinates
\beq
x^0\equiv c\, t,\,x^1,\,x^2\in{\mathbb R} \quad \text{and}\quad x^3 \in [0,L]\;.
\eeq
Now take, over this cylindrical manifold $\mathsf{M}^\prime$,
the chiral gauge field theory with
group $G$ and left-handed fermion representation $R_\text{\,left}$ given by:
\beq \label{SO10+Nfam=3}
\big(\,G \,; R_\text{\,left}\,\big) =
\big(\, SO(10)\, ; \mathbf{16}+\mathbf{16}+\mathbf{16}\,\big)\;,
\eeq
which incorporates the \SM~with three families of quarks and leptons
\cite{Zee82}. Moreover, let the fermions have periodic \bcs~in
$x^3$, \ie, a  periodic spin structure over $S^1$,
as indicated by the subscript $\text{PSS}$ in \eqref{R3S1}.

Then, for the theory as defined, quantum effects necessarily give
\DS{CPT} violation \cite{K00}, with a typical mass scale
\beq \label{msmall}
\msmall_\text{anom} \equiv \frac{\alpha_G \, \hbar}{L \,c}
  \approx
  2\times 10^{-35}\: {\rm eV}/c^2 \, \left(\,\frac{\alpha_G}{1/100}\, \right)\,
  \left(\,  \frac{10^{10}\,{\rm lyr}}{L}\,  \right),
\eeq
where $\alpha_G \equiv g^2/(4\pi)$ is defined in terms of the dimensionless
$SO(10)$ gauge coupling constant $g$ and $L$ is the size of the compact
dimension (here, taken as the size of the visible universe; see below).
As mentioned above, this phenomenon has been called a ``\DS{CPT} anomaly,''
the reason being that the \DS{CPT} invariance of the classical theory is broken
by quantum effects ($m_\text{anom} \propto \hbar \,$).

A heuristic argument for the existence of a \DS{CPT} anomaly
in theory \eqref{R3S1}--\eqref{SO10+Nfam=3} with appropriate
gauge field configurations runs as follows
\cite{K00,K98}:
\begin{itemize}
\vspace*{-0.00\baselineskip}
\item the periodic spin structure of the compact space dimension,
      with coordinate $x^3 \in [0,L]$,
      allows for momentum component $p_3 =0$ in a separable Dirac
      operator;\vspace*{-0.25\baselineskip}
\item a single four-dimensional chiral fermion with $p_3 =0$ corresponds
      to a single massless
      Dirac fermion in three dimensions;\vspace*{-0.25\baselineskip}
\item a single massless Dirac fermion in three dimensions is known to have
      a ``parity anomaly,'' provided gauge invariance is
      maintained \cite{R84,ADM85,CL89};\vspace*{-0.25\baselineskip}
\item this three-dimensional ``parity'' violation corresponds to \DS{T}
      violation in the original four-dimensional theory, which, in turn,
      leads  to \DS{CPT} violation.
\vspace*{-0.00\baselineskip}
\end{itemize}
Further discussion of this particular case will be postponed till
Sec.~\ref{Perturbative4D}. Here, we continue with some general remarks.

The heuristics of the previous paragraph suggests that the
\DS{CPT} anomaly also occurs for the $SO(10)$ theory \eqref{SO10+Nfam=3}
over $\mathbb{R} \times S^2  \times S^1_\text{\,PSS}$ or
$\mathbb{R} \times S^1\times S^1\times S^1_\text{\,PSS}$,
but not over $\mathbb{R} \times S^3$, where the Dirac operator is
nonseparable and the space manifold $S^3$ simply connected.
However, even over $\mathbb{R}^3 \times S^1_\text{\,PSS}\,$,
the \DS{CPT} anomaly does not occur for standard quantum
electrodynamics \cite{JauchRohrlich}, the vector-like gauge theory of
photons and electrons with $G=U(1)$ and $R_\text{\,left}= (1) + (-1)$.
Hence, \emph{both} nontrivial topology and parity violation are needed
for the \DS{CPT} anomaly.

Regarding the role of topology, the \DS{CPT} anomaly resem\-bles
the Casimir effect, with the local properties of the vacuum depending on
the \bcs~\cite{DeWitt75,Milton2001}.
Note that the actual topology of our universe is unknown \cite{LR99},
but theoretically there may be some constraints (cf. Ref.~\cite{CW95}).
Interestingly, the modification of the local physics due to the \DS{CPT} anomaly
would allow, in principle, for an indirect observation of the global spacetime
structure (see Sec.~\ref{MCSmodel+phenomenology}).

Clearly, it is important to
be sure of this surprising effect and to understand the mechanism better.
In the next section, we, therefore, turn to a relatively simple theory,
Abelian chiral gauge theory in two spacetime dimensions.
From now on, we put $\hbar$ $=$ $c$ $=$ $1$, except when stated otherwise.

\section{Exact result in two dimensions}
\label{Exact2D}

Consider chiral $U(1)$ gauge theory over the
flat torus $T^2$ $\equiv$ $S^1 \times S^1$,
with trivial zweibeins $e_\mu^a (x)= \delta_\mu^a$ and
Euclidean metric $g_{\mu\nu}(x) \equiv e_\mu^a (x)\,e_\nu^b (x)\, \delta_{ab}
=\delta_{\mu\nu}$, for diagonal matrix
$(\delta_{\mu\nu}) \equiv\mathrm{diag}(1,1)$.
In order to be specific, take the gauge-invariant theory with
five left-handed fermions of charges $(q_f)=(1, 1, 1, 1, -2)$
or $R_\text{\,left} = 4\times (1) +  1\times (-2)$.
Furthermore, impose doubly-periodic
boundary conditions on the fermions. The corresponding spin structure
will be denoted $\text{PP}$ and the specific theory $1111\overline{2}$.

The effective action $\Gamma\,[a]$ for the
$U(1)$ gauge field $a_\mu(x)$ is defined by the functional integral
\beqa  \label{pathint}
\exp\big(-\Gamma^{1111\overline{2}}_\text{PP}\,[a] \big) &=&
\int \prod_{f=1}^{5}
\left( {\mathcal D}\bar{\psi}_{Rf}{\mathcal D}\psi_{Lf}\right)_\text{PP}\,
\no\\
&& \times \,\exp\Big(-\sum_{f=1}^{5}\,\action_\text{\,Weyl}^{\,T^2}
\left[\bar{\psi}_{Rf},\psi_{Lf},q_f\,a\right]\Big)
\eeqa
and is known exactly \cite{IN99}.
In fact, the effective action is given in terms of Riemann
theta functions (see Appendix~\ref{Effective action 2D}).

It can now be checked explicitly that the \DS{CPT} transformation,
\beq \label{aCPTtransform}
a_\mu(x) \, \rightarrow \, a_\mu^\mathsf{CPT}(x) \equiv \,- a_\mu(-x)\,,
\eeq
does \emph{not} leave the effective action invariant  \cite{KN01}:
\beq \label{2dCPTanomaly}
\Gamma^{1111\overline{2}}_\text{PP}\,\big[\,a\,\big]  \,
\rightarrow \,
\Gamma^{1111\overline{2}}_\text{PP}\,\big[\,a^\mathsf{CPT}\,\big]= \,
\Gamma^{1111\overline{2}}_\text{PP}\,\big[\,a\,\big]
+ \,\pi \ii\, \pmod{2\pi \ii\,}\;.
\eeq
This result, which can also be understood heuristically (see
Appendix~\ref{Effective action 2D}),
shows unambiguously the existence of a \DS{CPT} anomaly in this
particular two-dimensional chiral $U(1)$ gauge theory. The crucial
ingredients are the dou\-bly-periodic ($\text{PP}$) \bcs~and
the odd number (here, five) of Weyl fermions.

\section{Nonperturbative result in four dimensions}
\label{Nonperturbative4D}

For two spacetime dimensions,
we have obtained in the previous section an exact result
for the effective action and established the precise form of the
\DS{CPT} anomaly, at least for appropriate \bcs.
In four dimensions, it is, of course, not possible to
calculate the effective action exactly.
Still, we can establish the \emph{existence} of the \DS{CPT} anomaly
by a careful consideration of the fermion measure.
This will be done nonperturbatively by use of a particular
lattice regularization of an Abelian \cgth.

Consider, then, the chiral $U(1)$ gauge theory
consisting of a single  gauge boson and sixteen left-handed
fermions with $U(1)$ charges $q_f$, for $f=1, \ldots , 16$.
Specifically,  the  gauge group and left-handed fermion representation
(\ie, the set of left-handed charges $q_f$) are given by:
\begin{subequations}
\beqa
G  &=& U(1),
\label{U1}\\[2mm]
R_\text{\,left}&=& 6\times (1/3) +  3\times (-4/3) + 3\times (2/3) +
        2\times (-1) + 1\times (2) \nonumber\\[1mm] & &  + 1\times (0)\; .
\label{U1charges}
\eeqa
\end{subequations}
This particular chiral $U(1)$ gauge theory can be embedded in
the $SU(2)\times U(1)$ theory relevant to the \SM~with
$U(1)$ hypercharge $Y\equiv 2\,Q-2\,T_3$; see, \eg, Ref.~\cite{Zee82}.
The further embedding in the ``safe'' $SO(10)$ group \cite{GeorgiGlashow1972}
explains that the perturbative  gauge anomalies cancel out for the chiral
$U(1)$ gauge theory considered: $\sum_f (q_f)^3 =0$
according to Eq.~\eqref{U1charges}.

Also take  a finite volume in Euclidean spacetime,
\begin{eqnarray}
V = L' \times L' \times L'  \times L\; , \label{volume}
\end{eqnarray}
and introduce a regular hypercubic lattice,
\begin{eqnarray}
L' = N'\, a\; , \quad L = N\, a\; ,\qquad N',N \in \mathbb{N}\; , \label{lengths}
\end{eqnarray}
with lattice spacing $a$
[not to be confused with the Abelian gauge field $a_\mu(x)$ in the continuum].
The lattice sites have coordinates
\begin{eqnarray}
(x_1,x_2,x_3,x_4) \equiv (\vx, x_4) = (\vn \, a, n_4\, a)\; ,
 \label{n1234}
\end{eqnarray}
for integers $n_1,n_2,n_3 \in [0,N'\,]$ and $n_4 \in [0,N\,]$.

The spinor fields $\psi_f$, with flavor index
$f=1,\ldots, 16$,
reside at the lattice sites and the vector field
$U_\mu(x) \in U(1)$ is associated with the directed link between site
$x$ and its nearest neighbor in the $\mu$--direction
(that is, between sites $x$ and $x+\widehat{\mu}\,$).
The \bcs~are taken to be periodic in $x_4$:
\begin{eqnarray}
\psi_f (\vx, L)= \psi_f(\vx, 0)\; ,\quad
U_\mu (\vx, L) = U_\mu (\vx, 0)\; , \label{pbcs}
\end{eqnarray}
mixed in $x_1$:
\begin{eqnarray}
\psi_f (L', x_2, x_3,x_4)&=& -\,\psi_f(0, x_2,x_3,x_4)  \: ,\nonumber\\[1mm]
U_\mu (L', x_2, x_3,x_4) &=& +\, U_\mu (0, x_2, x_3,x_4)\: ,
\label{apbc}
\end{eqnarray}
and similarly mixed in $x_2$ and $x_3$.

The specific \clgth~used has three main ingredients:
\begin{itemize}\vspace*{-0.00\baselineskip}
\item Ginsparg--Wilson fermions \cite{GW82};\vspace*{-0.25\baselineskip}
\item Neuberger's explicit lattice Dirac operator \cite{N98};\vspace*{-0.25\baselineskip}
\item L{\"u}scher's chiral constraints \cite{L99}.
\vspace*{-0.00\baselineskip}\end{itemize}
Technical details for the present setup can be found
in Ref.~\cite{KlinkhamerSchimmel}.

The theory is now well-defined and the
Euclidean effective gauge field action $\Gamma[U]$
can, in principle, be calculated by integrating out
the fermions ($U$ denotes the set of link variables).
The goal is to establish the following inequality
for at least one set of link variables:
\beq\label{GammaU-CPTanomaly}
\Gamma\big[\, U\, \big] \neq \Gamma \big[\, U^\mathsf{CPT}\, \big]\,,
\eeq
with $U^\mathsf{CPT}$ the set of \DS{CPT}--transformed link variables.

The result
\eqref{GammaU-CPTanomaly} has been obtained in Ref.~\cite{KlinkhamerSchimmel}
for an arbitrary odd number $N$ of links in the periodic
direction and for arbitrary values of the lattice spacing $a$.
As the continuum limit $a \to 0$ is not needed, the result is nonperturbative.
See Appendix~\ref{CPT anomaly on a 4D  lattice} for a sketch of the proof.

Most importantly, the \emph{origin} of the \DS{CPT} anomaly has been
identified \cite{KlinkhamerSchimmel} as an ambiguity in
the choice of basis vectors needed to define the fermion integration
measure, just as for Fujikawa's derivation \cite{F80,Nakahara90}
of the Abelian chiral anomaly (Adler--Bell--Jackiw triangle anomaly).

\section{Perturbative results in four dimensions}
\label{Perturbative4D}

In this section, we return to the spacetime continuum and consider
the four-dimensional chiral gauge theory of Sec.~\ref{Example}, with
\beq
G =SO(10)\,, \quad R_\text{\,left} = N_\mathrm{fam} \times (\mathbf{16})\,.
\label{SO10theoryGandRL}
\eeq
Two four-dimensional manifolds, called $\mathsf{M}^{\prime}$
and $\mathsf{M}^{\prime\prime}$, will be discussed explicitly.
From now on, the metric will have Lorentzian signature, with
spacetime indices running over $0$, $1$, $2$, $3$.

\subsection{Cylindrical manifold}
\label{Cylindrical manifold}

In this subsection, we take as a prototype of nontrivial
large-scale topology the cylindrical manifold discussed earlier.
Specifically, consider the chiral gauge theory \eqref{SO10theoryGandRL} over
\beq
\mathsf{M}^{\prime} =\mathbb{R}^3 \times S^1_\text{\,PSS}\,,
\quad
e^a_\mu (x)= \delta^a_\mu\,,
\quad
(\eta_{ab})\equiv\mathrm{diag}(1,-1,-1,-1) \,,
\label{SO10theoryManifoldM'}
\eeq
where PSS stands for periodic spin structure
with respect to the circle coordinate (denoted $x^3$ below)
and the metric is the standard Minkowski metric,
$g_{\mu\nu}(x) \equiv e_\mu^a (x)\,e_\nu^b (x)\, \eta_{ab} =\eta_{\mu\nu}$.

As mentioned before, the
four-dimensional effective action  $\Gamma[A]$,
for  $A$ $\in$ $\mathsf{so}(10)$, is  not known exactly.
But the crucial term has been identified perturbatively
for an appropriate class of gauge fields (indicated by a prime),
which has $A^\prime_3=0$ and $x^3$--independent fields in the
remaining three directions.
The effective action then contains the following term \cite{K00}:
\beq \label{CSlike}
\Gamma^{\,\mathsf{M}^{\prime}}[A'\,] \supseteq -
   \int_{{\mathbb R}^3} \rmd x^0 \rmd x^1 \rmd x^2\, \int_{0}^{L} \rmd x^3
   \; \frac{n \,\pi}{L}
   \: \omega_\text{\,CS}\big[ A'_0(x),A'_1(x),A'_2(x)\,\big] \,,
\eeq
with an integer $n$ and the standard \CS~density \cite{Nakahara90}
\beq \label{omegaCS}
 \omega_\text{\,CS} [ A_0,A_1,A_2\,] \equiv
       \frac{1}{16 \,\pi^2} \;\,\epsilon^{3\kappa\lambda\mu}\; \mathrm{tr}\,
       \big( F_{\kappa\lambda}\, A_\mu -
       (2/3)\, A_\kappa A_\lambda A_\mu \big),
\eeq
in terms of the \YM~field strength  \cite{YM54}
\beq
F_{\kappa\lambda} \equiv
\partial_\kappa A_\lambda - \partial_\lambda A_\kappa
+ A_\kappa A_\lambda - A_\lambda A_\kappa  \;,
\label{YMfieldstrength}
\eeq
where the fields and their derivatives are evaluated at the same
spacetime point.
Here, the gauge field takes values in the Lie algebra,
$A_\mu (x)$ $\equiv$ $A_\mu^a(x) \,T^a$ for $T^a \in \mathsf{so}(10)$
with normalization $\mathrm{tr}\, (T^a T^b)$ $=$ $(-1/2)\, \delta^{ab}$,
and $\epsilon^{\kappa\lambda\mu\nu}$ is the completely antisymmetric
Levi-Civita symbol with $\epsilon^{1230}= 1$.
The indices $\kappa,\lambda,\mu$ in  (\ref{omegaCS})
effectively run over $0, 1, 2$,
but the gauge fields may depend on \emph{all} coordinates:
$x^0$, $x^1$, $x^2$, \emph{and} $x^3$.
The  term (\ref{CSlike}) for general gauge fields is called
``\CS-like,'' because a genuine topological \CS~term exists only in an
odd number of dimensions \cite{Nakahara90}.

For gauge fields vanishing at infinity, replacing
$\epsilon^{3\kappa\lambda\mu}$ in the integrand of \eqref{CSlike}
by $\partial_\nu x^3\,\epsilon^{\nu \kappa\lambda\mu}$
and integrating by parts gives
the following manifestly gauge-invariant effective action:
\beq\label{CSlike-gaugeinvariant}
\Gamma^{\,\mathsf{M}^{\prime}}[A\,] =
 \int_{\mathsf{M}^{\prime}} \rmd^4 x  \;
 \frac{n}{32\pi}\,\frac{x^3}{L}\;\mathrm{tr}\,\bigl(
 \epsilon^{\kappa\lambda\mu\nu} \, F_{\kappa\lambda}(x)\,F_{\mu\nu}(x)\bigr)
 + \cdots\;,
\eeq
where the prime on $A$ has been dropped  and other terms, possibly nonlocal ones,
are contained in the ellipsis.

At this point, we can make three basic observations.
First, the local term \eqref{CSlike-gaugeinvariant},
with an explicit factor $x^3$ in the integrand, is clearly Lorentz noninvariant
and \DS{CPT} odd, in contrast to the \YM~action \cite{YM54},
\beq \label{SYM}
\hspace*{-0.75em}
\action_\text{\,YM}^{\,\mathsf{M}^{\prime}} =
     \int_{\mathsf{M}^{\prime}} \rmd^4 x \;
     \frac{1}{2\,g^2}\;
     \mathrm{tr}\,\bigl( \eta^{\kappa\mu}\,\eta^{\lambda\nu}\,
     F_{\kappa\lambda}(x)\,F_{\mu\nu}(x) \bigr).
\eeq
More precisely, the Lorentz and \DS{CPT} transformations considered are active
transformations on gauge fields of local support, as discussed in Sec. IV of Ref.
\cite{KN01} for the two-dimensional theory.
In physical terms, the wave propagation from the action
(\ref{SYM}) is essentially isotropic, whereas the term
\eqref{CSlike-gaugeinvariant} makes the propagation
anisotropic (see Sec.~\ref{MCS photons in flat spacetime}).

Second, the integer $n$ in the effective action term \eqref{CSlike-gaugeinvariant}
is a remnant of the ultraviolet regularization:
\beq  \label{n}
n \equiv \sum_{f=1}^{N_\mathrm{fam}} \;\big(2\, k_{0f}+1\big)\;,
\quad k_{0f} \in {\mathbb Z}\;.
\eeq
Since the sum of an odd number of odd numbers is odd, one has
$n \neq 0$ for  $N_\mathrm{fam}=3$
and the anomalous term \eqref{CSlike-gaugeinvariant} is necessarily present
in the effective action of the theory introduced in Sec.~\ref{Example}.

For $N_\mathrm{fam}=3$, the regularization of Ref.~\cite{K00} gives
minimally
\beq \label{nsimplest}
n = (1-1+1)\, \Lambda_0 \, /  \, |\Lambda_0| = \pm \,1 \;,
\eeq
with $\Lambda_0$ an ultraviolet Pauli--Villars
cutoff for the $x^3$--independent modes of the
fermionic fields contributing to the effective action.
[See Appendix~B of Ref.~\cite{KlinkhamerSchimmel} for a derivation of the odd
integers $2k_{0f}+1$ in Eq.~\eqref{n} from the lattice regularization.]
The effective action term \eqref{CSlike-gaugeinvariant} has, therefore,
a rather weak dependence on the small-scale
structure of the theory, as shown by the factor $\Lambda_0/|\Lambda_0|$
in  (\ref{nsimplest}).
This weak dependence on the ultraviolet cutoff has first been observed
in the so-called ``parity'' anomaly of three-dimensional gauge theories
\cite{R84,ADM85,CL89},
which underlies the four-dimensional \DS{CPT} anomaly as discussed
in Sec.~\ref{Example}.

Third, the $SO(10)$ theory \eqref{SO10theoryGandRL} for $N_\mathrm{fam}=3$
has \emph{three\,} identical irreps (irreducible representations)
and the \DS{CPT} anomaly \emph{must} occur
[the integer $n$ from (\ref{n}) is odd and therefore nonzero].
For the $SU(3) \times SU(2) \times U(1)$ \SM~with $N_\mathrm{fam}=3$,
the \DS{CPT} anomaly \emph{may or may not} occur,
depending on the ultraviolet regularization.
The reason is that the Standard--Model irreps come in \emph{even} number
(for example, four left-handed isodoublets per family),
so that the integer $n$ is not guaranteed to
be nonzero [$n$ is even and may or may not differ from zero];
see Sec. 5 of Ref.~\cite{K00} for details.
Note that the particular lattice gauge theory of  Sec.~\ref{Nonperturbative4D}
has all fermions regularized identically, so that the anomalous
terms do not cancel.

This concludes our discussion of the \DS{CPT} anomaly over cylindrical manifolds.
Section~\ref{MCSmodel+phenomenology}
considers certain phenomenological consequences,
whereas the next subsection
studies the anomalous effects from a different type of manifold.

\subsection{Punctured manifold}
\label{Punctured manifold}

In this subsection, we take as a prototype of nontrivial small-scale topology
the following ``punctured'' three-dimensional manifold:
\beq
\mathsf{M}^{\prime\prime}_3
= \mathbb{R} \times \big(\mathbb{R}^2 \setminus \{0\}\big)
= \mathbb{R}^3 \setminus \mathbb{R}\,.
\label{M''}
\eeq
The considered three-space may be said to have a linear ``defect,''
just as a type--II superconductor can have a single vortex line
(magnetic flux tube); cf. Ref.~\cite{Nakahara90}.
Furthermore, introduce cylindrical coordinates $(\rho,\phi, z)$ over
$\mathsf{M}^{\prime\prime}_3$,
\beq
\big(x^1,\,x^2,\,x^3\big) = \big(\rho \cos \phi,\,  \rho \sin\phi,\, z\big)\,,
\label{cylindricalcoord}
\eeq
with the $z$--axis at the position of the line puncture (linear defect)
and coordinate domains
$\rho \in (0,\infty)$,  $\phi \in [0,2\pi]$, and  $z \in (-\infty, \infty)$.

The corresponding four-dimensional spacetime
manifold $\mathsf{M}^{\prime\prime}=\mathbb{R} \times \mathsf{M}^{\prime\prime}_3$
is orientable and has flat metric $(\eta_{\mu\nu})=\mathrm{diag}(1,-1,-1,-1)$,
but with nontrivial vierbeins.
The particular theory considered in this subsection is, in fact,  given by
\eqref{SO10theoryGandRL} and
\newcommand{\fourmat} [1]{\left( \begin{array}{cccc} #1 \end{array} \right)}
\beq
\mathsf{M}^{\prime\prime} =\mathbb{R} \times \mathsf{M}^{\prime\prime}_\text{3,\,PSS}\,,
\quad \big(e^a_\mu (x)\big) =
\fourmat{ 1 & 0 & 0 & 0\\
          0 & \cos\phi & -\sin\phi & 0\\
          0 & \sin\phi & \cos\phi & 0\\
          0 & 0 & 0 & 1  }
\,,
\label{SO10theoryManifoldM''}
\eeq
with the vierbeins shown in matrix notation.
Again, PSS stands for periodic spin structure, but now
with respect to the coordinate $\phi$.
One particular class of noncontractible loops in $\mathsf{M}^{\prime\prime}_3$
consists of circles with fixed values of $\rho$ and $z$
(these circles are noncontractible because of the
line removed from $\mathbb{R}^3$, which happens to coincide with the
$z$ axis of the coordinates used).

For our purpose, it suffices to establish
the \DS{CPT} anomaly for one particular class of gauge fields.
Take the four-dimensional gauge fields over $\mathsf{M}^{\prime\prime}$
to be independent of $\phi$ and without
 component in the direction of $\phi$.
These fields will be indicated by a double prime  in the following.
The anomalous contribution to the effective action is then
found to be given by \cite{KlinkhamerRuppPRD2004}
\begin{equation}
\Gamma^{\,\mathsf{M}^{\prime\prime}}[A''\,] \supseteq
\int_{\mathsf{M}^{\prime\prime}} \dint{^4x} \,
\frac{n}{32\pi}\;\frac{\phi(x)}{2\pi} \;\;
\mathrm{tr}\,\bigl(
\epsilon^{\kappa\lambda\mu\nu} \,F''_{\kappa\lambda}(x)\, F''_{\mu\nu}(x)
\bigr)\,,
\label{nonAbeliananomM''}
\end{equation}
where $\phi(x)$ denotes the azimuthal angle from \eqref{cylindricalcoord},
measured with respect to the linear defect of $\mathsf{M}^{\prime\prime}_3$.
The long-range anomalous effects occur already for an infinitely thin
linear defect, which is not the case for standard electromagnetic
propagation effects. Furthermore, the anomalous term \eqref{nonAbeliananomM''}
from nontrivial small-scale topology
(noncontractible loops with arbitrarily small lengths)
has the same structure as
\eqref{CSlike-gaugeinvariant} from nontrivial large-scale topology
(noncontractible loops with lengths equal to or larger than $L$).
A result similar to \eqref{nonAbeliananomM''} has been obtained
heuristically \cite{KlinkhamerRuppPRD2004} for a space manifold
$\mathbb{R}^3$ with two points identified, which is a simplified
version of a permanent static ``wormhole'' \cite{W57,V96}.

The general structure of the anomalous term (\ref{nonAbeliananomM''}) for
an arbitrary flat manifold $\mathsf{M}$
with a single puncture (or wormhole) has the following form:
\begin{equation}
\Gamma^{\,\mathsf{M}}[A\,] =
 \int_{\mathbb{R}^4} \dx f_\mathsf{M}(x;A]\;\;\mathrm{tr}\,\bigl(
\epsilon^{\kappa\lambda\mu\nu} \,
 F_{\kappa\lambda}(x)\, F_{\mu\nu}(x)\bigr) + \cdots\,,
\label{anom3}
\end{equation}
where $F_{\mu\nu}$ stands for the
Yang--Mills field strength \eqref{YMfieldstrength} and the
integration domain has been  extended to $\mathbb{R}^4$,
which is possible for smooth enough gauge fields $A_\mu(x)$.
The factor $f_\mathsf{M}(x;A]$ is both a function of the spacetime
coordinates $x^\mu$
and a gauge-invariant functional of the gauge field $A_\mu(x)$.
This functional dependence of $f_\mathsf{M}$ involves, most likely,
the gauge field holonomies.
But the functional $f_\mathsf{M}(x;A]$ is not known in general.

This concludes our brief discussion of the \DS{CPT} anomaly over a manifold
with a single puncture. The calculation with two or more punctures (or wormholes)
is, however, difficult and a simple phenomenological model will be introduced
in  Sec.~\ref{Random-coupling model}.

\section{Maxwell--Chern--Simons model and phenomenology}
\label{MCSmodel+phenomenology}

\subsection{MCS action and microcausality}
\label{MCS action}

Starting from the four-dimensional continuum theory of
Sec.~\ref{Cylindrical manifold},
we consider the electromagnetic $U(1)$ gauge field $a_\mu(x)$
embedded in the $SO(10)$ gauge field $A_\mu(x)$.
Also, we extend the cylindrical manifold $\mathsf{M}^{\prime}$ to
Minkowski spacetime ${\mathbb R}^4$ with
metric $(\eta_{\mu\nu})=\mathrm{diag}(1,-1,-1,-1)$.
In effect, we take the double limit $L \to \infty$ and $n \to \infty$
of \eqref{CSlike}  and \eqref{SYM}, with constant ratio $n/L$.
For most of this section, we will suppress the explicit spacetime
dependence of the fields.

For electromagnetic fields $a_\mu$ of local support and after appropriate
rescaling, the following local terms can be expected to be present
in the effective action:
\beqa \label{S-MCS}  %%\tfr{1}{4}
\action_\text{\,MCS}  &=&
\action_\text{\,M}  + \action_\text{\,CS}  \;,  \\[0.25cm]
\action_\text{\,M}  &=&
 \int_{{\mathbb R}^4} \rmd^4 x \;
      \big(  - (1/4)\, \eta^{\kappa\mu}\,\eta^{\lambda\nu} \,
              f_{\kappa\lambda}\,f_{\mu\nu} \big)\;,
\label{Mterm}\\[0.25cm]
\action_\text{\,CS}  &=&
 \int_{{\mathbb R}^4} \rmd^4 x  \;
      \big(    + (1/4)\,\msmall\,
                \epsilon^{3\kappa\lambda\mu} \, f_{\kappa\lambda}\,a_\mu \big)\;,
\label{CSterm}
\eeqa
with Maxwell field strength
\beq
f_{\mu\nu} \equiv \partial_\mu a_\nu - \partial_\nu a_\mu
\label{Maxwellfieldstrength}
\eeq
and \CS~mass parameter $\msmall \geq 0$ [in terms of the previous
parameters: $\msmall \sim \alpha\, n/L$, with
fine-structure constant  $\alpha\equiv e^2/(4\pi)$ and $e \sim g$].

The  Maxwell--\CS~(MCS) model \emph{per se} has been studied
before,  in particular,
by the authors of Refs.~\cite{CFJ90,AndrianovSoldati98,CK98}.
The action \eqref{S-MCS} is gauge
invariant, provided the electric and magnetic fields in
$f_{\kappa\lambda}$ vanish fast enough at infinity.
The gauge invariance of the \CS-like term \eqref{CSterm}
makes clear that the parameter $\msmall$ is not sim\-ply the mass
of the photon \cite{GN71}, it affects the propagation in a different way
(see Sec.~\ref{MCS photons in flat spacetime}).

On the other hand, there is known to be a close relation
\cite{J57,SW64,J65,BLOT90} between
\DS{CPT} invariance and microcausality, \ie, commutativity of
local observables with spacelike separations. The question is then
whether or not causality holds in the \DS{CPT}--violating MCS model.
Remarkably, microcausality (locality) can be established
also in the particular MCS model considered \cite{AK01npb}.
The commutation relations are given in Appendix~\ref{Microcausality in 4D MCS}.

The topics discussed in the remainder of this section include
the propagation properties of MCS photons and their
interactions with conventional electrons and gravitational fields.
Note that, even though certain results are obtained for
classical waves, we will speak freely about ``photons,''  assuming that the
complete quantization procedure can be performed successfully \cite{AK03,CK97}.

\subsection{MCS photons in flat spacetime}
\label{MCS photons in flat spacetime}

The propagation of electromagnetic waves
in the Maxwell--\CS~(MCS) model (\ref{S-MCS})
makes clear that \DS{C} and \DS{P} are conserved, but \DS{T} \emph{not}.
An example is provided by the behavior of pulses of circularly polarized light,
as will be shown in this subsection.

The dispersion law for plane electromagnetic waves in the MCS model
is given by \cite{AK01npb,CFJ90,AndrianovSoldati98,CK98}:
\beq \label{eq:disprel}
\omega_{\,\pm}^2 \equiv k_1^2 + k_2^2 +
                    \bigl( q_3 \pm \msmall/2   \bigr)^2 \; ,
                    \quad q_3 \equiv \sqrt{k_3^2 + \msmall^2 /4} \;,
\eeq
where the suffix $\pm$ labels the two different modes
(denoted $\plus$ and $\minus$, respectively). The phase and group
velocities are readily calculated from this dispersion law,
\beq \label{eq:vphvg}
\vv_{\rm ph}^{\, \pm} \equiv \bigl(k_1,k_2,k_3\bigr) \,
                                \frac{\omega_{\, \pm}}{|\vk\,|^2}    \;, \quad
\vv_{\rm g}^{\, \pm}  \equiv
\left( \frac{\partial}{\partial k_1}\,, \frac{\partial}{\partial k_2}\,,
       \frac{\partial}{\partial k_3}\,\right)\omega_{\,\pm}\; .
\eeq
The magnitudes of the group velocities turn out to be given
by (recall $c \equiv 1$):
\beq \label{eq:absvg2}
|\vv_{\rm g}^{\, \pm} (k_1,k_2,k_3)|^2 = \frac
{k_1^2 + k_2^2 +  \bigl(q_3 \pm \msmall/2 \bigr)^2 \,k_3^2 /q_3^2 }
{k_1^2 + k_2^2 +  \bigl(q_3 \pm \msmall/2 \bigr)^2 } \leq 1 \;,
\eeq
with equality for $m=0$  or for a $\minus$--mode having $k_3=0$
(recall $\msmall \geq 0$).

For our purpose, it is necessary to obtain the explicit polarizations
of the electric and magnetic fields (see Refs.~\cite{AK03,KaufholdK05} for
further details). As long as the propagation of the plane wave is
not exactly along the $x^3$ axis, the radiative electric field can be
expanded as follows ($\Re$ denotes taking the real part):
\beqa \label{E}
\vE_{\,\pm} (\vx,t) &=&
\Re\,\Big(
c^{\,\pm}_1\; \big(\, \widehat{\ve}_3 -
(\,\widehat{\ve}_3 \cdot \widehat{\vk}\,)\:\widehat{\vk}\,\big) +
c^{\,\pm}_2\; \big(\, \widehat{\ve}_3 \times \widehat{\vk} \,\big) +
c^{\,\pm}_3\; \widehat{\vk} \,  \Big) \no\\[2mm]
&&
\times \exp \left[\, \ii\, (\vk \cdot \vx - \omega_{\,\pm} \,t) \,\right] ,
\eeqa
with unit vector $\widehat{\ve}_3$ in the preferred $x^3$--direction,
unit vector $\widehat{\vk}$ corresponding to the wave vector $\vk\,$,
polar angle $\theta$ of the wave vector
(so that $k_3 \equiv \vk \cdot \widehat{\ve}_3 = |\vk|\,\cos\theta$),
and complex coefficients $c^{\,\pm}_1$, $c^{\,\pm}_2$, and $c^{\,\pm}_3$
(at this point, the overall normalization is arbitrary).
The vacuum MCS field equations then give
the following polarization coefficients for the two modes:
\beq \label{Epols}
\left(\begin{array}{c}
c^{\,\pm}_1 \\[1mm]
c^{\,\pm}_2 \\[1mm]
c^{\,\pm}_3
\end{array}\right) =
\left(\begin{array}{c}
\cos\theta \left(\sqrt{\cos^2\theta +
\mu_{\,\pm}^2 \sin^4\theta} \,\pm\,  \mu_{\,\pm} \,\sin^2\theta\right)^{-1}
\\ \pm \,\ii\,  \\[1mm]
\mp \,2\, \mu_{\,\pm} \, \sin^2\theta
\end{array}\,\right)\;,
\eeq
with $\mu_{\,\pm} \equiv m/(2\,\omega_{\,\pm}) \geq 0$ for
positive frequencies $\omega_{\,\pm}$ from Eq.~(\ref{eq:disprel}).
The corresponding magnetic field is
\beq
\vB_{\,\pm} = \bigl(\,\vk \,\times \vE_{\,\pm}\, \bigr)/\,\omega_{\,\pm}\;.
\eeq

\begin{figure}[t]
\begin{center}
\includegraphics[width=7.5cm]{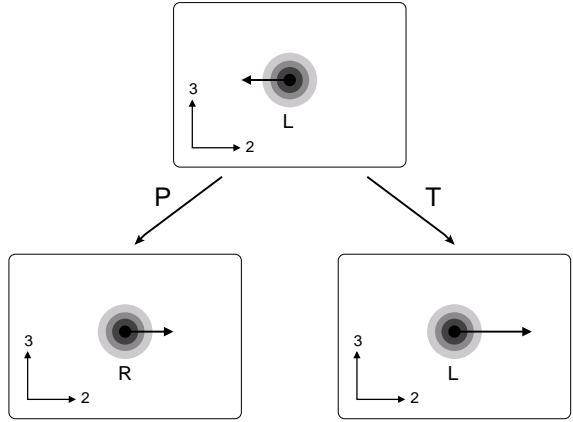}
\end{center}
\caption{Sketch of the behavior of a left-handed wave packet in the \MCS~model
  (\ref{S-MCS}) under the time reversal (\DS{T}) and parity (\DS{P}) transformations.
  [The charge conjugation (\DS{C}) transformation acts trivially.]
  The nonzero energy density of the pulse is indicated by the shaded area and
  the arrow
  shows the group velocity approximately along a ``standard'' direction with
  coordinate  $x^2$,
  for the case of a ``preferred'' direction with coordinate $x^3$.
  The magnitude of the group velocity changes under \DS{T},
  but not under \DS{C} or \DS{P}.  Hence, the physics is \DS{CPT}--noninvariant.
  In addition, the vacuum is seen to be optically active,
  with left- and right-handed light pulses traveling to the right at different speeds
  (the same holds for pulses traveling to the left).}
\label{pulses}
\end{figure}

As long as
the $ \mu_{\,\pm} \sin^2\theta$ terms in (\ref{Epols}) are negligible
compared to $|\cos\theta|$, the transverse electric field
consists of the standard circular polarization modes (see below).
For the opposite case, $|\cos\theta|$ negligible compared to $ \mu_{\,\pm} \sin^2\theta$,
the transverse polarization ($c^{\,\pm}_1$, $c^{\,\pm}_2$) becomes effectively
linear.

Now consider the propagation of light pulses close to the $x^2$ axis.
For $k_1=0$ and $0 < \msmall  \ll |k_3| \ll |k_2|$, in particular,
we can identify the $\plus/\minus$--modes of the dispersion law (\ref{eq:disprel})
with left- and right-handed circularly polarized modes
($L$ and $R\,$; cf. Ref.~\cite{J75}),
depending on the sign of $k_3 \equiv |\vk|\,\cos\theta$.
From Eqs. (\ref{E}) and (\ref{Epols}), one obtains that
$\plus/\minus$ corresponds to $R/L$ for $k_3>0$ and to $L/R$ for $k_3<0$.

With these identifications, Eq.~(\ref{eq:absvg2}) gives the following relations
for the group velocities of pulses of circularly polarized light
($\msmall  \ll |k_3| \ll |k_2|$):
\begin{subequations}
\beqa
|\,\vv^{\,L}_\mathrm{g}(0,k_2,k_3)| &=&  |\,\vv^{\,R}_\mathrm{g}(0,-k_2,-k_3)|\;,
\label{vgP}\\[2mm]
|\,\vv^{\,L}_\mathrm{g}(0,k_2,k_3)| &\neq& |\,\vv^{\,L}_\mathrm{g}(0,-k_2,-k_3)|\;,
\label{vgT}
\eeqa
\end{subequations}
provided $\msmall \neq 0$.
Recall, at this point, that the time-reversal operator \DS{T} reverses the
direction of the wave vector and leaves the helicity unchanged,
whereas the parity-reflection operator \DS{P} flips both  the wave vector and the helicity.
Equality (\ref{vgP}) is, therefore, consistent with  parity invariance, while
inequality (\ref{vgT}) implies
time-reversal noninvariance for this concrete physical situation
(see Fig.~\ref{pulses}).

The velocities \eqref{eq:vphvg}--\eqref{eq:absvg2}
show that the vacuum has become optically active  (see also Fig.~\ref{pulses}).
In particular,
left- and right-handed monochromatic plane waves travel at different speeds
\cite{CFJ90}. (This effect has also been noticed
by the authors of Ref.~\cite{HuangSikivie1985}
in the context of axionic domain walls.)
In the following two subsections, we discuss two ``applications''
of MCS optical activity or birefringence.

\subsection{Cosmic microwave background}
\label{CMB}

\begin{figure}[t]
\begin{center}
\includegraphics[width=10cm]{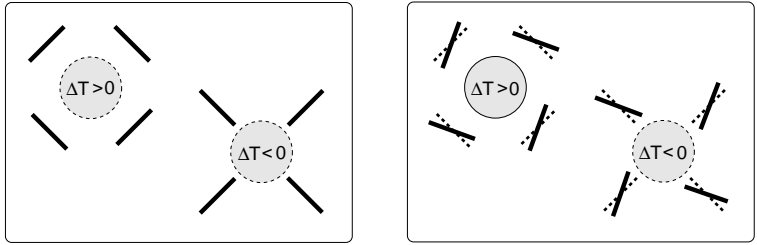}
\end{center}
\caption{Sketch of the linear polarization pattern (indicated by heavy bars)
      around cosmic-microwave-background hot-spots ($\Delta T>0$) and
      cold-spots ($\Delta T<0$), generated by scalar perturbations of
      the metric. The left panel is in
      a ``standard'' direction with Cartesian coordinate $x^1$ or $x^2$.
      The right panel is in the ``preferred''
      direction with coordinate $x^3$ and displays the optical activity
      of the \MCS~model (\ref{S-MCS}) considered.  In fact,
      for a patch of sky in a particular direction along the  $x^3$ axis
      (shown in the right panel), the linear
      polarization pattern is
      rotated by a very small amount in the counterclockwise direction.
      For a patch of sky in the opposite direction (not shown),
      the rotation of the linear polarization is in the clockwise direction.}
\label{CMBpol}
\end{figure}

In the previous subsection, we have seen that
the  MCS vacuum is optically active.
As mentioned in Ref. \cite{K00}, this may, in principle,
lead to observable effects of the \DS{CPT} anomaly
in the cosmic microwave background (CMB): the polarization pattern
around hot-spots and cold-spots is modified due to the
action of the \CS-like term (\ref{CSterm}) on the  electromagnetic
waves traveling between the last-scattering
surface (redshift $z$ $\sim$ $10^3$) and the detector ($z$ $=$ $0$).
Figure \ref{CMBpol} gives a sketch of this cosmic
birefringence effect, which can be looked for by  ESA's Planck Surveyor
and next-generation satellite experiments (perhaps CMBPOL).
See Ref. \cite{HW97} for a pedagogical review of the expected CMB polarization
and Ref. \cite{L98}
for further details on the possible signatures of cosmic birefringence
from a spacelike \CS~vector
(a timelike \CS~vector was considered in Ref.~\cite{LWK99}).

It is important to realize that the optical activity from the \DS{CPT} anomaly, as
illustrated by Fig.~\ref{CMBpol}, is essentially frequency independent,
in contrast to the quantum-gravity effects suggested by the authors
of, for example, Refs.~\cite{GP99,AMU02}.
Quantum-gravity effects on the photon propagation can generally be expected
to become more and more important as the photon
energy increases towards $E_\text{\,Planck}
\equiv (\hbar\, c^5/G)^{1/2}\approx 1.2 \times 10^{19}\,\text{GeV}$.
The potential \DS{CPT}--anomaly effect at the relatively low CMB photon
energies ($\hbar\,\omega\sim 10^{-4}\,\text{eV}$) is, therefore,
quite remarkable. Indeed,
the weak ultraviolet-cutoff dependence of the \DS{CPT} anomaly
has already been commented on a few lines below Eq.~(\ref{nsimplest}).

\subsection{Big bang vs. big crunch}
\label{BBvsBC}

In this subsection, we turn to an entirely different application
of MCS photons, namely as an ingredient of a \emph{Gedankenexperiment}.
The problem addressed, the arrow of time, is one of the most profound of modern
physics and we refer to the clear discussion given by Penrose \cite{P79};
further references can be found in, \eg, Ref.~\cite{Zeh99}.

\begin{figure}
    \begin{center}
      \includegraphics[width=0.7\textwidth]{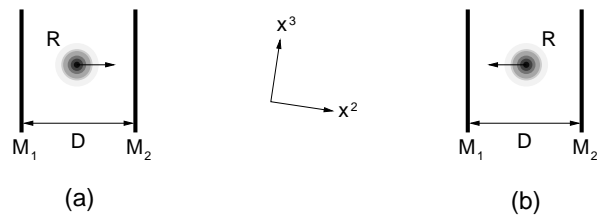}
    \end{center}
\caption{(a) Sketch of clock $C$, with a single pulse of circularly
polarized light reflecting between two parallel mirrors, $M_1$ and $M_2$,
at a fixed distance $D$. Shown is the time at which the clock is started,
with a right-handed ($R$) light pulse moving towards the right
(\protect\ie, in the direction of increasing $x^2$).
(b) Sketch of clock $C^\prime$, which has all motions reversed compared to
clock $C$. Clock $C^\prime$ starts with a right-handed light pulse moving
towards the left.}
\label{lightclocks-construction}
  \end{figure}
\begin{figure}
    \begin{center}
      \includegraphics[width=0.7\textwidth]{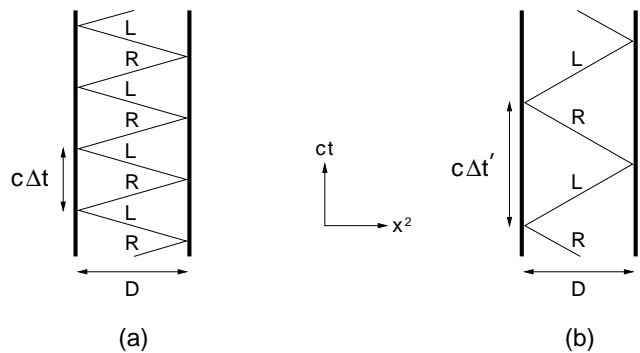}
    \end{center}
\caption{(a) Spacetime diagram of clock $C$ in the MCS model \eqref{S-MCS},
with ticks $\Delta t$ between the successive reflections of the light
pulse. The slight offset in the $x^3$--direction, as indicated by
Fig.~\ref{lightclocks-construction}, is not shown here.
 (b) Spacetime diagram of clock $C^\prime$, with ticks $\Delta t^\prime$.}
\label{lightclocks-spacetimediagram}
\end{figure}

After examining the various time-asym\-me\-tries present at the
macroscopic level, Penrose asked the basic question:
\emph{``what special geometric structure did the big bang possess that
distinguishes it from the time-reverse of the generic
singularities of collapse---and why?''}

He then proposed a particular condition (the vanishing of the
Weyl curvature tensor) to hold at any \emph{initial}
singularity. Whatever the precise condition may be, the crucial point is
that this condition would \emph{not} hold for
\emph{final} singularities.
This implies that the unknown physics responsible for the
``initial singularity'' necessarily involves \DS{T}, \DS{PT}, \DS{CT},
and \DS{CPT} violation; see Sec. 12.4 of Ref.~\cite{P79}.

But Penrose did not make a concrete proposal
for the \emph{physical mechanism} of this \DS{T} and \DS{CPT} noninvariance.
In Ref.~\cite{K02}, the possible relevance of the \DS{CPT}
anomaly was suggested, which does not
involve gravitation directly but does depend on the global structure
(topology) of space.

Consider the light clock $C$ of Figs. \ref{lightclocks-construction}a
and \ref{lightclocks-spacetimediagram}a.
The decisive point, now, is that the time-reversed copy $C'$ of
Figs. \ref{lightclocks-construction}b and \ref{lightclocks-spacetimediagram}b
runs differently in the effective MCS model \eqref{S-MCS},
as discussed in Sec.~\ref{MCS photons in flat spacetime}.

\begin{figure}
    \begin{center}
      \includegraphics[width=0.4\textwidth]{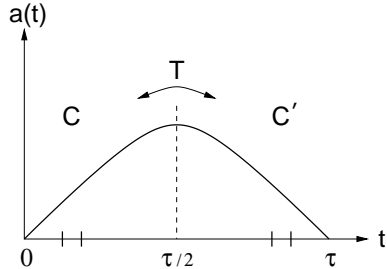}
    \end{center}
\caption{Clocks $C$ and $C^\prime$ in a Kantowski--Sachs universe
    with expansion factor $a$ as a function of cosmic time $t$
    and re-collapse time $\tau$.
    The background metric is invariant under time reversal (\DS{T}) but
    the clock $C$ made from MCS photons not.}
\label{KSuniverse}
\end{figure}

More fundamentally, consider the $SO(10)$ \cgth~\eqref{SO10+Nfam=3}
in a homogeneous  Kantowski--Sachs universe
\cite{CW95,KantowskiSachs1966,Collins1977}
with spacetime topology ${\mathbb R} \times S^2  \times S^1_\text{\,PSS}$,
which re-collapses after a period of expansion and
has anomalous \DS{CPT} violation.
The clock $C$ near the big bang and the time-reversed copy of clock $C$
(\ie, clock $C^\prime$)  near the big crunch then give a
\emph{different number of ticks} over an equal time interval as defined by
a standard clock (or by the expansion and contraction of the universe).
The setup is sketched in Fig.~\ref{KSuniverse}.
(See also Ref.~\cite{AharonyNeeman70} for a related discussion of a
$K^0$--beam with hypothetical \DS{CPT} violation in a re-collapsing universe.)

Therefore, the physics near the initial singularity and the physics near the
final singularity could be different, as demonstrated by this
\emph{Gedankenexperiment} with MCS photons in a  Kantowski--Sachs
universe \cite{K02}.
Of course, the potential effect discussed gives only a ``direction
in time'' and the main dynamics of the big-bang singularity still needs
to be explained. In a way, the situation would be analogous to spontaneous
magnetization in ferromagnets, where a small impurity or boundary effect
determines the direction in space of the  magnetization in the domain considered
but the dynamics is really driven by the spin interactions.

\subsection{Decay processes in modified QED}
\label{Decay processes}

%%FRK: \xi was written as \widehat\vz=\widehat\vec\zeta in \cite{KaufholdK05}

In this subsection, we consider new two- and three-particle
decay processes \cite{KaufholdK05} in the Maxwell--Chern--Simons
model with conventional electrons added. Using the inverse Minkowski metric
$(\eta^{\mu\nu})\equiv\mathrm{diag}(1,-1,-1,-1)$ to raise indices
and to define $\gamma^\mu\gamma^\nu + \gamma^\nu\gamma^\mu= \eta^{\mu\nu}$,
the relevant action of this particular modification of
quantum electrodynamics (QED) is  given by
\begin{equation}\label{eq:mod-QED}
\action_\text{\,modified\;QED}=
\action_{\,\text{MCS},\;\xi^\mu\xi_\mu=-1\;,\;\xi^0=0}\:
 +\action_{\,\text{D}}\; \text,
\end{equation}
with the Maxwell--Chern--Simons (MCS) terms
\beq
\action_{\,\text{MCS},\;\xi^\mu} =
\int_{{\mathbb R}^4} \rmd^4 x \;
      \big(\,  - (1/4)\, f_{\mu\nu}\,f^{\mu\nu}
              + (1/4)\;\, m\,\epsilon_{\mu\nu\rho\sigma}\,
                \xi^{\,\mu}\,a^\nu\, f^{\rho\sigma}\, \big)\text,
\label{eq:MCSterm}
\eeq
and the standard Dirac term \cite{JauchRohrlich}
\begin{equation}\label{eq:Dterm}
\action_{\,\text{D}}= \int_{\mathbb{R}^4} \rmd^4 x \; \bar \psi \,
\big(\,\ii\,\gamma^\mu\partial_\mu - M - e \,\gamma^\mu a_\mu\big)\,\psi\text,
\end{equation}
where the electron from field $\psi$ has charge $e$ and mass $M>m/2>0$.
Remark that the normalized (dimensionless) \CS~vector $\xi^\mu$ has been
taken to be purely spacelike in \eqref{eq:mod-QED} and that
the corresponding spatial vector $\boldsymbol{\xi}$ was
previously taken to point in the $x^3$--direction,
as shown by \eqref{CSterm}. Note also that $\xi^\mu$
was written as $\widehat\zeta^\mu$ in Ref.~\cite{KaufholdK05}.
The two polarization modes of the MCS photon are again denoted
$\plus/\minus$, corresponding to the $+/-$ sign in the dispersion
law \eqref{eq:disprel}.

First, we discuss
the Cherenkov process $e^-\to\minus\, e^-$, which occurs
already at tree level and is allowed for any three-momentum
$\vq$ of the  electron, provided $\vq$ has a nonzero component in the
$\boldsymbol{\xi}$--direction.
The process $e^-\to\plus\, e^-$ is not allowed kinematically.
(See, \eg,  Refs.~\cite{ColemanGlashow1997,Jacobson-etal2005} for a
general discussion of vacuum Cherenkov radiation
and Ref.~\cite{LehnertPotting} for a discussion in the
context of the MCS model.)

\begin{figure}
\begin{center}
      \includegraphics[width=0.6\textwidth]{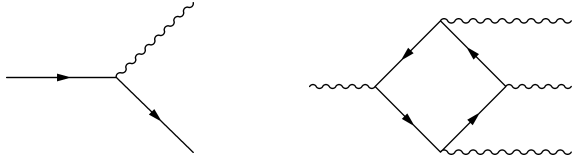}
\end{center}
\caption{Feynman diagrams contributing to vacuum Cherenkov radiation
    (left panel) and photon triple-splitting (right panel) in the
    Maxwell--Chern--Simons model with conventional Dirac fields added.}
\label{VacuumCherenkovPhotonSplitting}
\end{figure}

The tree-level amplitude $A$ for $e^-\to\minus\, e^-$
follows directly from the QED interaction \eqref{eq:Dterm},
\begin{equation}
A = \bar u(q-k)\, \bar\epsilon_\mu(k)\, (- e \gamma^\mu)\, u(q)\text,
\label{eq:cherenkov-amplitude}
\end{equation}
with $u$ the incoming and $\bar u$ the outgoing spinor and
$\bar\epsilon_\mu$ the conjugate polarization vector of the MCS photon.
The corresponding Feynman diagram is shown in the left panel of
Fig.~\ref{VacuumCherenkovPhotonSplitting}.
(The Feynman rules of standard QED are given in, for example,
Ref.~\cite{JauchRohrlich}.)

An analytic calculation gives the following Cherenkov decay width
\cite{KaufholdK05}:
\begin{equation}\label{CherenkovDecayWidth}
\Gamma(\vq)=\frac{1}{2\,\sqrt{|\vq|^2+M^2}}\; \gamma(\qp)\text,
\end{equation}
with decay parameter $\gamma$ as a function of the parallel
momentum $\qp \equiv \vq \cdot \boldsymbol{\xi}$:
\beqa\label{gammacherenkov}
\hspace*{-14mm}&&\gamma(\qp) =\;
\frac{\alpha}{16\,\sqrt{\qp^2+M^2}}\,\Big[\,8 m^2 \,  |\qp|
- 2 m \, \big(4\,  |\qp| -k_\text{max} \big )\, \sqrt{m^2 + 4 k_\text{max}^2}
\no\\[2mm]
\hspace*{-14mm}&&
-4  \left(m^2 + 4M^2\right) k_\text{max}
+ m \big(m^2 + 8 M^2 + 16\, \qp^2\,\big)\,
  \text{arcsinh}\Big(\, \frac{2\, k_\text{max}}{m} \,\Big) \Big]\text,
\eeqa
for  fine-structure constant  $\alpha\equiv e^2/(4\pi)$
and maximum parallel photon momentum $k_\text{max}$ defined by
\begin{equation}\label{eq:kmax}
k_{\text{max}}(\qp) \equiv
\fracnew{2 m \, |\qp|\,\left(m +2\, \sqrt{\qp^2+M^2}\,\right)}
{m^2+4M^2+4 m\, \sqrt{\qp^2+M^2}}\; \geq \, 0\,\text.
\end{equation}
For $0\leq\qpabs<M$, the result \eqref{gammacherenkov} can be expanded
in $m/M$,
\begin{equation}\label{gammacherenkov-lowE}
\gamma(\qp)= (4/3)\,\alpha\; m\,\qpabs^3/ M^2
 + \BigO\left( \alpha\, m^2\, \qpabs^3 / |M|^3 \,\right)\text,
\end{equation}
while, for $\qpabs\gg M$, an expansion in $m/\qpabs$ and $M/\qpabs$ gives
\begin{equation}\label{gammacherenkov-highE}
\gamma(\qp)=
\alpha \,m \, \qpabs \, \big(\ln(\qpabs/m)+2\ln 2-3/4\,\big)+ \cdots\,\text,
\end{equation}
where the ellipsis stands for subdominant terms.
Hence, the decay parameter of the electron grows approximately linearly
with the momentum component in the preferred direction, but is
suppressed by one power of $m$.
For $|\vq| \to \infty$ and fixed angle $\theta \ne \pi/2$
between $\vq$ and $\vxi$,
the  decay rate \eqref{CherenkovDecayWidth} behaves as follows:
\begin{equation}\label{CherenkovDecayWidth-highE}
\Gamma(\vq) \sim  (1/2)
\:\alpha \,m \, |\cos\theta|\;\ln \big( |\vq|/m \big)\text,
\end{equation}
where the definition
$\qp \equiv \vq \cdot \boldsymbol{\xi} \equiv
|\vq| \cos\theta$ has been used and
only the leading term in $|\vq|$ has been shown.

Next, we discuss photon triple-splitting in the purely spacelike
MCS model \eqref{eq:MCSterm}, which was first considered in
Ref.~\cite{AK03} and then generalized in Ref.~\cite{KaufholdK05}.
There are eight decay channels, corresponding to all possible
combinations of $\plus$--modes and $\minus$--modes.
It can be shown that the following three
channels are allowed for generic initial three-momentum~$\vq$:
$\plus\to\minus\minus\minus$, $\plus\to\plus\minus\minus$, and
$\minus\to\minus\minus\minus$, whereas the five others are
kinematically forbidden. For special momentum $\vq \perp \boldsymbol{\xi}$,
only the decay channel  $\plus\to\minus\minus\minus$ is available.

The implication would be that, with suitable interactions,
all MCS photons are  generally unstable against splitting.
The exception would be for the lower-dimensional subset
of  $\minus$--modes with three-momenta orthogonal to $\boldsymbol{\xi}$.

The interaction is now taken to be the
Euler--Heisenberg interaction and the photonic action considered reads
\begin{align}\label{eq:MCSEHaction}
\action_\text{photon}&=
\action_{\text{MCS},\;\xi^\mu\xi_\mu=-1\;,\;\xi^0=0}
+\action_\text{EH}\text,
\end{align}
consisting of the quadratic MCS terms \eqref{eq:MCSterm},
for purely spacelike background four-vector $\xi^\mu$,
and the quartic Euler--Heisenberg term
\begin{equation}
\hspace*{-1mm}
\action_\text{EH}=
\frac{2\alpha^2}{45 M^4} \int_{\mathbb{R}^4} \rmd^4 x\, \left[\,\big(
\,(1/2) \, f_{\mu\nu}f^{\mu\nu}\big)^2
+7\, \big(\,(1/8) \, \epsilon_{\mu\nu\rho\sigma}
f^{\mu\nu}f^{\rho\sigma}\big)^2\,\right]\text,
\label{eq:EHterm}
\end{equation}
with fine-structure constant $\alpha\equiv e^2/(4\pi)$ and  electron mass $M$.
For modified QED with action \eqref{eq:mod-QED},
the Euler--Heisenberg term arises from  the low-energy limit
of the  one-loop electron contribution
to the effective gauge field action  \cite{JauchRohrlich};
see also the  right panel of Fig.~\ref{VacuumCherenkovPhotonSplitting}.

The decay width of photon triple-splitting in model \eqref{eq:MCSEHaction}
is then given by \cite{KaufholdK05}:
\begin{equation}\label{eq:decay-width-photon-splitting}
\Gamma(\vq)= \frac{1}{2 \,\omega(\vq)}\;\gamma(\qp)\text,
\end{equation}
with the following behavior of the decay parameter for $\qpabs \gg m$:
\begin{equation}\label{eq:gamma-photon-splitting}
\gamma(\qp)\sim  c \; \alpha^4 \; m^5\; \qpabs^5/M^8 \text.
\end{equation}
The numerical constant $c$ in \eqref{eq:gamma-photon-splitting}
depends on the decay channel
($\plus\to\minus\minus\minus$, $\plus\to\plus\minus\minus$, or
$\minus\to\minus\minus\minus$)
and ranges between $1.23 \times 10^{-10}$ and $2.33\times 10^{-10}$.

Finally, let us comment on the possible high-energy behavior of
photon triple-splitting in modified QED with action \eqref{eq:mod-QED},
as our calculation in model \eqref{eq:MCSEHaction}
was only valid for momenta less than
$M^2/m$, with an extra factor $M/m$ compared to the naive expectation
$M$ \cite{KaufholdK05}. Recall that, for  standard QED, the
$\BigO(\alpha^2)$ amplitude of a four-photon interaction
is known in principle \cite{JauchRohrlich}.

Consideration of the amplitude and phase space integral suggests the
following behavior for the decay parameter of the process
shown in the right panel of Fig.~\ref{VacuumCherenkovPhotonSplitting}:
\begin{equation}\label{eq:gamma-conjecture}
\gamma\;\big|_{\;\qpabs \gg M^2/m}
\stackrel{?}{\sim}\; c_\infty\,\alpha^4\,m\,\qpabs  \,\text,
\end{equation}
neglecting logarithms of $\qpabs$.
Combined with the ``low-energy'' result \eqref{eq:gamma-photon-splitting},
this would imply that the effect of Lorentz breaking
continues to grow with energy. At ultra-high energies,
the decay rate \eqref{eq:decay-width-photon-splitting}
would then approach a direction-dependent constant (up to logarithms).
A similar behavior has been seen for
vacuum Cherenkov radiation in \eqref{CherenkovDecayWidth-highE}.

\subsection{MCS photons in curved spacetime backgrounds}
\label{MCS photons in curved spacetime}

%%FRK: $\xi^\mu$ was written as $-\zeta^\mu$ in \cite{KantK05}

The MCS model \eqref{S-MCS} can also be coupled to gravity. One
possibility for the coupling is given by the following
generalized action \cite{KantK05,Kostelecky2004}:
\begin{eqnarray}
\hspace*{-1.5em}
\action&=& \action^\text{grav.}_\text{EH} +
            \action^\text{gen.}_\text{MCS}  +  \cdots \,,
\label{S-general}\\[2mm]
\hspace*{-1.5em}
\action^\text{grav.}_\text{EH}&=&\int \rmd^4 x \; e\medspace
R/(16\pi G)    \,,
\label{S-EH}\\[2mm]
\hspace*{-1.5em}
\action^\text{gen.}_\text{MCS}&=&\int \rmd^4 x \;
\left(- \frac{1}{4}\,e\,
   g^{\kappa\mu}g^{\lambda\nu}\,f_{\kappa\lambda}\,f_{\mu\nu} +
   \frac{1}{4}\,m  \;\xi_a e_\kappa^{\;a}\,
\epsilon^{\kappa\lambda\mu\nu}  f_{\lambda\mu}\,a_\nu \right)\,,
\label{S-MCSgeneral}
\end{eqnarray}
for the case of a Cartan connection
$\Gamma^\lambda_{\;\mu\nu}=\Gamma^\lambda_{\;\nu\mu}$
(\ie, a torsion-free theory \cite{Wald}),
so that the standard definition \eqref{Maxwellfieldstrength} of the field
strength $f_{\mu\nu}$ still holds.
Note that $\xi_a$ was written as $-\zeta_a$ in Ref.~\cite{KantK05}.
In addition, $g_{\mu\nu}(x)$ is the metric with signature $(+\,-\,-\,-\,)$,
$e_\kappa^{\;a}(x)$ the vierbeins with
$e(x)\equiv \det e_\kappa^{\;a}(x)$, $R$  the Ricci  curvature
scalar which enters the Einstein--Hilbert action \eqref{S-EH} with
a coupling proportional to
the inverse of Newton's constant $G$,
and $\epsilon^{\kappa\lambda\mu\nu}$ the Levi--Civita tensor density.

The combined action from Eqs.~\eqref{S-EH} and \eqref{S-MCSgeneral}
is, however, not satisfactory \cite{Kostelecky2004} and
further contributions are needed, hence the ellipsis
in Eq.~\eqref{S-general}.
For the moment, we only consider the light-propagation effects from the
MCS action \eqref{S-MCSgeneral} in given spacetime backgrounds.

The condition $\partial_b \, \xi^a =0$ holds for the flat MCS model
and the covariant generalization $D_\mu \xi_\nu=0$ might seem natural.
But this condition imposes
strong restrictions on the curvature of the spacetime \cite{Kostelecky2004}
and it may be better to demand only closure \cite{CFJ90},
\begin{equation}\label{zetaa}
D_\mu\xi_\nu-D_\nu \xi_\mu=\partial_\mu \xi_\nu-\partial_\nu
\xi_\mu=0.
\end{equation}
This last requirement
ensures, at least, the gauge invariance of action \eqref{S-MCSgeneral}.
Furthermore, we assume that the norm of $\xi^\mu$ is constant,
$\xi_\mu\xi^\mu=-1$,
in order to simplify the calculations.

The geometrical-optics approximation of the MCS model
\eqref{S-MCSgeneral} in a curved spacetime background
has been studied in Ref.~\cite{KantK05}.
The main result there is the derivation of a modified geodesic equation,
starting from the equation of motion of the gauge field,
\begin{equation}\label{gaugefieldeqn}
e\,D_\mu f^{\mu\nu}= (1/2)\,m\,
\xi_\kappa\,\epsilon^{\kappa\nu\rho\sigma}\, f_{\rho\sigma}\,.
\end{equation}
A plane-wave \emph{Ansatz},
\begin{equation}
a^\mu (x)=C^\mu(x)\, \exp[\,\ii\, S(x)]\,,
\end{equation}
gives then in the Lorentz gauge $D_\mu a^\mu=0$:
\beq
e \left( D_\mu S \right)  \left( D^\mu S \right)  C^\nu =
\ii\, m\,\xi _\kappa \epsilon^{\kappa \nu \rho \sigma}\,
\left( D_\rho S \right) C_\sigma \,,
\quad
D_\mu D^\mu S = 0,
\label{eikonaleqs}
\eeq
where derivatives of the complex amplitudes $C^\mu$
and  a term involving the Ricci tensor have been neglected
(the typical length scale of $a_\mu$
is assumed to be much smaller than the length scale of the
spacetime background). The equality signs in (\ref{eikonaleqs})
are, therefore, only valid in the geometrical-optics limit.
As usual, the wave vector is defined to be normal to surfaces of equal phase,
\begin{equation}\label{kmu}
k_\mu\equiv D_\mu S.
\end{equation}
See, \eg, Refs.~\cite{Wald,BornWolf}
for further discussion of the geometrical-optics approximation.

Equations~\eqref{eikonaleqs} give essentially the same dispersion law as in
flat spacetime. There exist, again, two inequivalent modes,
one with mass gap and the other without,
\begin{equation}\label{DPR resolved}
k^\mu k_\mu=m^2/2\pm  \sqrt{m^4/4+m^2\,(\xi^\mu k_\mu)^2}\;.
\end{equation}
For  $k^\mu k_\mu \ne 0$,
the following ``modified wave vector'' can be defined \cite{KantK05}:
\begin{equation}
\widetilde{k}^\mu\equiv
k^\mu- m^2 \big( k^\mu+\xi^\mu\xi^\nu k_\nu \big)/(2 k^\rho k_\rho),
\label{ktilde}
\end{equation}
which has constant norm, $\widetilde{k}^\mu\widetilde{k}_\mu= m^2/4>0$.
The crucial observation, now, is that
this modified wave vector obeys a geodesic-like equation,
\begin{equation}\label{ktildegeodesic}
\widetilde{k}^\mu D_\mu\widetilde{k}_\lambda=0,
\end{equation}
whereas $k^\mu$ generally does not.

In the flat case, $\widetilde{k}^\mu$ corresponds to the group velocity,
which is also the velocity of energy transport
\cite{Brillouin}. Hence, $\widetilde{k}^\mu$ must, in general,
be tangent to the geodesic that  describes the path of a ``light ray.''
Because the norm of $\widetilde{k}^\mu$ is positive,
Eq.~\eqref{ktildegeodesic} describes \emph{timelike} geodesics instead of the
standard null geodesics for Maxwell light rays.
The vector $k^\mu$ in the Maxwell--Chern--Simons model, defined by \eqref{kmu},
no longer points to the direction in which the wave propagates,
but the vector $\widetilde{k}^\mu$, defined by \eqref{ktilde}, does.

The propagation of MCS light rays in Schwarzschild and
Ro\-bert\-son--Wal\-ker  backgrounds \cite{Wald} can now be calculated.
In particular, for the Schwarzschild metric with line element
\begin{equation}\label{Schwarzschild}
\rmd s^2=\left( 1-\frac{2G M}{r}\right)\rmd t^2-\left(
1-\frac{2G M}{r}\right)^{-1}\rmd r^2-r^2\rmd \theta^2-r^2\sin^2\theta\,\rmd \phi^2,
\end{equation}
two noteworthy results have been found \cite{KantK05}:
\begin{itemize}
\vspace*{-0.00\baselineskip}
\item
the existence of \emph{stable} circular orbits of MCS light rays
with radii larger than $6G M$,
whereas ``standard'' photons have only one unstable orbit
with radius $3G M$;\vspace*{-0.25\baselineskip}
\item
the possibility of \emph{different} gravitational redshifts of the two
MCS polarization modes.
\vspace*{-0.00\baselineskip}
\end{itemize}
Here, we only elaborate on the second result and
consider, for simplicity, the approximation of having a wave vector
$\vk$  parallel to the \CS~vector $\boldsymbol{\xi}$ at the two
points considered, $P_1$ and $P_2$.
Denoting the $\plus$--mode and $\minus$--mode
by subscripts `$+$' and `$-$' on $\omega$ and letting
$\omega_{\pm,j}$ refer to a static observer at point $P_j\,$,
the gravitational redshift is found to be given by:
\begin{equation}
\left. \frac{\omega_{\pm,1}-\omega_{\pm,2}}{\omega_{\pm,1}}\;\,
\right|^{\text{\,parallel}}_\text{\,MCS}
=\left(1 \mp \frac{m}{2\,\omega_{\pm,1}}\right)\,
\Delta_\text{standard}^\text{(Schwarz.)}\,,
\label{pararedshift-MCS}
\end{equation}
in terms of the result for standard photons,
\begin{equation}\label{pararedshift-standard}
\Delta_\text{standard}^\text{(Schwarz.)}
\equiv 1- \sqrt{(1-2G M/r_1)/(1-2G M/r_2)} \,.
\end{equation}
While the gravitational redshift of standard photons ($m\equiv 0$)
in a Schwarz\-schild background is the same for both polarization modes,
the redshift of ``parallel'' MCS photons differs by a relative
factor $mc^2/(\hbar\omega)$, with $\hbar$ and $c$
temporarily reinstated. A similar result holds for MCS photons in
a Ro\-bert\-son--Wal\-ker  background.

The unusual intrinsic properties of MCS photons thus lead
to interesting effects in curved spacetime backgrounds.
But the gravitational back-reaction of MCS photons remains a major
outstanding problem.

\section{Random-coupling model and photon propagation}
\label{Random-coupling model}

\subsection{Photon model and dispersion law}
\label{Photon model}

As mentioned in Sec.~\ref{Punctured manifold}, we are
faced with the difficulty of performing the anomaly calculation
already for two punctures (or other defects such as wormholes).
For this reason, we restrict ourselves to an Abelian gauge field
and simply introduce a ``random'' (time-independent) background field $g$
over $\mathbb{R}^4$ to mimic the anomalous
effects of a multiply connected (static) spacetime foam,
generalizing the result \eqref{nonAbeliananomM''}--\eqref{anom3}
of a single defect. The phenomenological model consists of
this frozen field $g(x)$ and a dynamical photon field $a_\mu(x)$,
both defined over the auxiliary manifold $\mathbb{R}^4$ with Minkowski metric
$\eta_{\mu\nu}$ of signature $(+\,-\,-\,-\,)$. In this section,
the spacetime dependence of the fields will be shown explicitly.

The photon model is then given by the action \cite{KlinkhamerRuppPRD2004}
\beq
\action_\text{\,photon}^{\,[g(x)]} =
  \int_{\mathbb{R}^4} \dx   \big( -(1/4) f_{\mu\nu}(x)  f^{\mu\nu}(x)
  -(1/4) \, g(x)\,  f_{\kappa\lambda}(x)  \widetilde f^{\kappa\lambda}(x)\big)\, ,
  \label{Gammaphoton}
\eeq
with Maxwell field strength  $f_{\mu\nu}(x)$ defined by
\eqref{Maxwellfieldstrength} and its dual by
$\widetilde f^{\kappa\lambda}(x)\equiv
(1/2)\;\epsilon^{\kappa\lambda\mu\nu}\, f_{\mu\nu}(x)$, for
\LC~symbol $\epsilon^{\kappa\lambda\mu\nu}$.
Note the important simplification in going from \eqref{anom3}
to \eqref{Gammaphoton}, where the gauge-field-independent
random coupling constant $g(x)$ makes the model action
quadratic in the photon field $a_\mu(x)$.
The additional term in the action density of \eqref{Gammaphoton}
can also be written in the form of an
Abelian Chern--Simons-like term, namely proportional to
$\partial_\mu g(x) \,\epsilon^{\mu\nu\rho\sigma} f_{\nu\rho}(x)\, a_\sigma(x)$.

Models of the type \eqref{Gammaphoton}
have been considered before, but only for
coupling constants $g(x)$ varying smooth\-ly over cosmological
scales; cf. Refs.~\cite{CFJ90,KLP02}.
Here, the assumed properties of the background field $g(x)$ are
very different \cite{KlinkhamerRuppPRD2004}:
\begin{itemize}
\vspace*{-0.00\baselineskip}
\item time independence, $g=g(\vx)$;\vspace*{-0.25\baselineskip}
\item weakness, $|g(\vx)| = \mathrm{O}(\alpha) \ll 1$;\vspace*{-0.25\baselineskip}
\item small-scale variation of $g(\vx)$  over length scales
      which are negligible compared to the wavelengths
      of the photon field $a_\mu(x)$;\vspace*{-0.25\baselineskip}
\item vanishing $g(\vx)$ average in the large-volume limit;\vspace*{-0.25\baselineskip}
\item finiteness, isotropy, and cutoff of the $g(\vx)$ autocorrelation
      function.
\vspace*{-0.00\baselineskip}
\end{itemize}
The modified Maxwell equation in the Lorentz gauge ($\partial_\nu a^\nu=0$)
now reads:
\begin{equation}
\Box \, a^\nu(x) = - \partial_\mu g(x)\, \widetilde f^{\mu\nu}(x)\,.
\label{photon_field_eq_pos}
\end{equation}
The dispersion law of the transverse modes can then be calculated by
expanding the solution to second order in $g$,
under the assumption that the power spectrum of $g(\vx)$
vanishes for momenta $|\vq|<q_\mathrm{low}$
and that the photons have momenta $|\vk| < q_\mathrm{low}/2\,$.

In the long-wavelength limit, the following dispersion law
of (transverse) photons is found \cite{KlinkhamerRuppPRD2004}:
\begin{equation}
\omega^2 = \left(1-A^2 \, \gamma_1\right)\, k^2
 -  A^2 \, \lgamma^2  \,k^4 + \mathrm{O}(k^6)\,,
\label{disp_foam}
\end{equation}
with simplified notation $k\equiv |\vk|\,$
and $g(\vx)$ amplitude $A \sim \alpha$.
The constants $\gamma_1$ and $\lgamma$ in \eqref{disp_foam}
are functionals of
the random couplings $g(\vx)$. Specifically, they are given by
\beq
\gamma_1 = \frac{\pi}{18\,A^2} \, C(0) \,, \quad
\lgamma^2 = \frac{2\pi}{15\,A^2} \int_0^\infty \dint{x} x\,C(x)\,,
\eeq
in terms of the isotropic autocorrelation function
$C(x)=\widehat{C}(\vx)$, for $x=|\vx|$, which has the general definition
\beq
\widehat{C}(\vx) \equiv \lim_{R\to \infty} \frac{1}{(4\pi/3) R^3}
\int\limits_{|\vy|<R} \rmd^3 y\;  g(\vy) \, g(\vy + \vx)\,.
\label{correl}
\eeq

The calculated dispersion law \eqref{disp_foam} is Lorentz noninvariant
($\omega^2 - c^2\,|\vk|^2$ $\ne$ $\text{constant}$)
but still \DS{CPT} invariant, even though
the original model action \eqref{Gammaphoton} also violates \DS{CPT}.
The explanation is that
the assumed randomness of $g(\vx)$ removes the anisotropies in the
long-wavelength limit. This modified dispersion law
can now be tested, in particular, by high-energy astrophysics.

\subsection{Experimental limits}
\label{Experimental limits}

In this subsection, we discuss a single ``gold-plated'' event:
an ultra-high-energy cosmic ray observed on October 15,
1991, at the Fly's Eye Air Shower Detector in Utah,
with energy $E\approx 3 \times 10^{11} \,\text{GeV}$ \cite{Bird-etal1995}.

For definiteness, assume an unmodified proton dispersion law
$E_p^2=k^2+m_p^2$ (recall $\hbar=c=1$)
and a modified photon dispersion law (\ref{disp_foam}).
The absence of Cherenkov-like processes $p \to p \gamma$
\cite{ColemanGlashow1997} for a proton energy of the order of
$E_p  \approx 3 \times 10^{11} \, \text{GeV}\,$ then
gives ``experimental'' limits \cite{KlinkhamerRuppPRD2005,GagnonMoore}:
\beq
\gamma_1 < \left(6\times 10^{-19}\,\right)\left(\alpha/A\right)^2, \quad
\lgamma  <  \left(1.0 \times 10^{-34}\,\text{cm}\right)\left(\alpha/A\right),
\eeq
with fine-structure constant $\alpha \approx 1/137$ inserted for $A$.

The basic astrophysical input behind these limits
has been reviewed in Ref.~\cite{KlinkhamerRuppNewAstRev2005},
which also discusses time-dispersion limits which are
less sharp but more direct.
The physical interpretation of these bounds in terms of the
structure of the underlying manifold is an open problem,
the work of Refs.~\cite{KlinkhamerRuppPRD2004,KlinkhamerRuppPRD2005}
being very preliminary.

\section{Conclusion}
\label{Conclusion}

The possible influence of spacetime topology on the local properties of
\qfth~has long been recognized (\eg, for the Casimir effect).
As discussed in the present contribution,
it now appears that nontrivial topology
may also lead to \DS{CPT} noninvariance for chiral gauge field theories such as
the \SM~with an odd number of families.
This holds even for flat spacetime manifolds, that is,  without gravity.

As to the physical origin of the \DS{CPT} anomaly, many questions remain
(the same can be said about chiral anomalies in general).
It is, however, clear that the
gauge-invariant second-quantized vacuum state plays a crucial role in
connecting the global spacetime structure to the local
physics \cite{K00,KlinkhamerSchimmel}. In a way,
this is also the case for the Casimir effect \cite{DeWitt75,Milton2001}.
New here is the interplay of parity violation (chiral fermions)
and gauge invariance.
Work on this issue is in progress (the most promising are perhaps small
lattice models), but progress is slow.\footnote{Another possible
source of \DS{CPT} violation may be a new type of quantum phase transition
in a fermionic quantum vacuum \cite{KV-qpt-jetpl,KV-emergent-ijmpa},
which, in the context of elementary particle physics,
could manifest itself via neutrino oscillations
\cite{K-neutrinos-jetpl,K-neutrinos-ijmpa,K-theta13-prd}.}

As to possible applications of the \DS{CPT} anomaly,
we have, first, considered non\-trivial \emph{large-scale} topology.
An example would be the flat spacetime manifold
$\mathsf{M}={\mathbb R} \times S^1 \times S^1 \times S^1_\text{\,PSS}\,$,
with time coordinate $x^0\equiv c\, t \in{\mathbb R}$
and PSS standing for periodic spin structure.
The anomaly may then give rise to new effects in photon physics, such as
vacuum bi\-re\-frin\-gence, photon triple-splitting,
and stable orbits of light around a nonrotating central mass.
Furthermore, we have discussed the potential role of
the \DS{CPT} anomaly as one ingredient for
the very special initial conditions of our universe, which
may be needed to explain the observed arrow of time.

Next, we have considered
a hypothetical \emph{small-scale} topology
of spacetime, which can also be probed by the \DS{CPT} anomaly.
From experimental results in cosmic-ray physics, it appears
possible to  obtain upper bounds on certain characteristic
length scales of a (static) spacetime foam.

But more important than these particular applications is the general idea:
\emph{spacetime topology affects the second-quantized vacuum
of \cgth~and the fundamental symmetries of the theory
(Lorentz and \DS{CPT} invariance),
which, in turn, provides a way to investigate certain properties of spacetime.}

\vspace*{.5\baselineskip}
The author would like to thank his collaborators of the last six years
for their valuable contributions,
the organizers of this conference for their hospitality,
and Gustavo C. Branco for providing the happy occasion.

\appendix

\section{Effective action of two-dimensional chiral $U(1)$ gauge theory}
\label{Effective action 2D}

The two-dimensional Euclidean action for a single one-component Weyl
field $\psi(x)$ of unit charge ($q=1$) over the particular torus $T^2$
with modulus $\tau=\ii\,$ is given by
\beq \label{eq:action}
\action_\text{\,Weyl}^{\,T^2}\left[\bar{\psi},\psi,a\right]
 =   - \int_{0}^{L} \rmd  x^1 \int_{0}^{L}\rmd  x^2 \;e\: \bar{\psi}\:
       e^\mu_a\,\widetilde{\sigma}^a \left(\partial _\mu +\ii\, a_\mu  \right)  \psi \;,
\eeq
with
\beq
   (\widetilde{\sigma}^1, \widetilde{\sigma}^2) = (1,\ii\, )\,, \;\;
   e^\mu_a = \delta^\mu_a\,\,, \;\; e \equiv \det\left( e_\mu^a\right) = 1 \,.
\eeq
The $U(1)$ gauge potential can be decomposed as follows:
\beq
a_\mu (x)= \epsilon_{\mu\nu}\, \delta^{\nu\rho}\,\partial _\rho \phi (x)  +
           2\pi h_\mu/L + \, \partial_\mu \chi (x)  \;,
\label{eq:Adecomposed}
\eeq
with $\phi(x)$ and $\chi(x)$  real periodic functions
and  $h_1$ and $h_2$ real constants. In this decomposition,
$\chi(x)$ corresponds to the gauge degree of freedom.
The related gauge transformations on the fermion fields are
\beq
\psi(x)       \rightarrow \exp[-\ii\, \chi(x)\,]\;\psi(x) \;, \quad
\bar{\psi}(x) \rightarrow \exp[+\ii\, \chi(x)\,]\;\bar{\psi}(x) \;.
\eeq
Next, impose doubly-periodic \bcs~on the fermions,
\beq
   \psi  (x^1 + L, x^2) =  \,\psi  (x^1,x^2)\;, \quad
   \psi  (x^1, x^2 + L) =  \,\psi  (x^1,x^2)\;.
\eeq
This spin structure will be denoted $\text{PP}$,
where $\text{P}$ stands for periodic \bcs.
(The other spin structures are $\text{AA}$, $\text{AP}$, and $\text{PA}$,
where $\text{A}$ stands for antiperiodic
\bcs. See, \eg, Ref.~\cite{AvisIsham79} for a general discussion
of how to deal with the different spin structures.)

The effective action $\Gamma\,[a]$ of the $(1111\overline{2})$--theory
from Sec.~\ref{Exact2D}, defined by the functional integral (\ref{pathint}),
is found to be given by \cite{IN99}:
\beq \label{GammaPP}
\exp\left(-\Gamma^{1111\overline{2}}_\text{PP}\,[a] \right) \equiv
D^{1111\overline{2}}_\text{PP}\,[a]
  = \left( D_\text{PP}\,[a]  \right)^4 \;
           \overline{\left( D_\text{PP}[2a] \right)}\;,
\eeq
in terms of the single chiral determinant
\beqa
\label{eq:DPP}
D_\text{PP}\,[a] &=&
\widehat{\vartheta} \big(h_1 +1/2, h_2 + 1/2 \big)\;
                 \exp\big(\ii\, \pi\,(h _ 1 - h _2)/2 \big)\;
\no\\[1mm]
&&
\times \exp\left(\frac{1}{4\pi}\int_{T^2} \rmd ^2 x
     \left( \phi \,\partial ^2 \phi + \ii\,  \phi \,\partial ^2
     \chi\right)\right) .
\eeqa
Here, the complex-valued function
\beqa
\label{eq:thetahat}
\widehat{\vartheta} (x,y) &\equiv& \exp\left(- \pi y^2 + \ii\,  \pi x y\right)
                                    \, \vartheta (x+\ii\,  y;\ii\,)/ \eta (\ii\,)\; ,
\;\; \text{for}\;\; x,y \in {\mathbb R} \;,
\eeqa
is defined
in terms of the Riemann theta function $\vartheta (z;\tau)$
and De\-de\-kind eta function $\eta (\tau)$, both for modulus $\tau=\ii\,$.
The bar on the \rhs~of Eq.~(\ref{GammaPP}) denotes complex conjugation.

The gauge invariance of the effective action (\ref{GammaPP})
can be readily verified.
In fact, the gauge degree of freedom $\chi(x)$ appears
only in the last exponential of Eq. (\ref{eq:DPP}),
namely in the term proportional to $\ii\,  \phi \,\partial ^2 \chi$,
and cancels out for the full expression (\ref{GammaPP})
since $4 \times 1^2$ $-$ $1 \times 2^2$ $=0$.
More work is needed
to show the invariance  under large gauge transformations,
$h_\mu$ $\rightarrow$ $h_\mu+n_\mu$ for $n_\mu \in {\mathbb Z}$.

The \DS{CPT} anomaly (\ref{2dCPTanomaly}) follows directly
from the $\vartheta$--function properties, as shown in Ref.~\cite{KN01}.
The relevant properties of $\vartheta (z;\tau)$
are its periodicity under $z$ $\rightarrow$ $z+1$ and
quasi-periodicity under $z$ $\rightarrow$ $z+\tau$, together with the
symmetry $\vartheta (-z;\tau)$ $=$ $\vartheta (z;\tau)$.
But the anomaly can also be understood heuristically from the
product of eigenvalues. For gauge fields (\ref{eq:Adecomposed}) with
$\phi(x)$ $=$ $\chi(x)$ $=$ $0$
and infinitesimal harmonic pieces $h_\mu$, one has, in fact,
\beq \label{eq:D11112linear}
D^{1111\overline{2}}_\text{PP}\,[h_1,h_2] =
\kappa\,(h_1 + \ii\,  h_2)^3\,(h_1^2 + h_2^2) + \BigO (h^7) \;,
\eeq
with a nonvanishing complex constant $\kappa$. Clearly, this expression
changes sign under the transformation
$h_\mu \rightarrow - h_\mu$, which corresponds to the \DS{CPT} transformation
(\ref{aCPTtransform}).

By choosing topologically nontrivial zweibeins $e_\mu^a(x)$
[still with a flat metric
$g_{\mu\nu}(x)\equiv  e_\mu^a(x)\;e_\nu^b(x)\;\delta_{ab}=\delta_{\mu\nu}\,$]
and including the spin connection term in the covariant derivative of the
fermionic action (\ref{eq:action}),
the \DS{CPT} anomaly can be moved to the spin structures AA, AP, and PA.
These topologically nontrivial zweibeins
correspond to the presence of spacetime torsion, which may be of
interest in itself. See Ref. \cite{KM01} for further details on the
possible  role of topologically nontrivial torsion.

\section{CPT anomaly on a four-dimensional lattice}
\label{CPT anomaly on a 4D  lattice}

In this appendix, we sketch the main steps for establishing the \DS{CPT}
anomaly on a four-dimensional lattice \cite{KlinkhamerSchimmel}.
The Euclidean chiral $U(1)$ gauge theory
considered has already  been defined in Sec.~\ref{Nonperturbative4D}.

First, restrict the gauge field configurations to those with
trivial link variables in the periodic direction ($\mu=4$) and
$x_4$--in\-de\-pen\-dent link variables in the other directions
($\mu=m=1,2,3$):
\begin{eqnarray}
U_4 (\vx, x_4) = \openone\; ,\quad U_m (\vx, x_4) = U_m (\vx) \; .
\label{config}
\end{eqnarray}
Next, introduce Fourier modes for the fermion field (single flavor)
\begin{eqnarray}
\psi(x) = \sum_n \psi_{n} (\vx) \: \exp\big(2\pi \ii\,  n x_4 /L\big)\; ,
\label{fermimod}
\end{eqnarray}
where the integer $n$ takes the values
\begin{subequations}
\begin{eqnarray}
-(N-1)/2 \leqslant &n& \leqslant (N-1)/2, \;\;\, \mbox{for odd $N$},
\label{nodd}\\[1mm]
-(N/2) +1 \leqslant &n& \leqslant (N/2), \;\,\,\qquad \mbox{for even $N$}.
 \label{neven}
\end{eqnarray}
\end{subequations}
Having made these choices, the integral for the effective action
\emph{factorizes}:
\beq
\hspace*{-0.25em}
\exp \big( -\Gamma[U] \big) =
K\, \prod_n \int \prod_{j} \mathrm{d} c_j^{(n)} \mathrm{d} \bar c_j^{(n)}
\,\exp \Big( - \sum_{j,k} \bar c_k^{(n)}\,
  M_{kj}^{(n)}[U]\, c_j^{(n)} \Big),
\label{integral}
\eeq
with constant $K$, Grassmann numbers $c_j^{(n)}$ and $\bar c_j^{(n)}$, and
matrices
\begin{eqnarray}
M_{kj}^{(n)} [U] \equiv a^3 \sum_{\vx} \bar v_k^{(n)} (\vx)\, a D^{(n)}[U] \,
v_j^{(n)} (\vx; U], \label{matrix}
\end{eqnarray}
where $D^{(n)}[U]$ is a three-dimensional Dirac operator.
The vectors $\bar v_k^{(n)}$  and $v_j^{(n)}$
build complete orthonormal bases of lattice
spinors satisfying  the appropriate chiral constraints.
Note that, in the present formalism \cite{GW82,N98,L99},
the left-handed basis vectors $v_j^{(n)}$ depend on the gauge-field
configuration $U$, as indicated on the \rhs~of \eqref{matrix}.

The \DS{CPT}--transformed link variables are:
\begin{eqnarray} \label{Utheta}
U^\theta_4 = U_4 = \openone\; ,\quad
U^\theta_m (\vx) \equiv U^\dagger_m (-\vx -a\,\widehat{\vm})\; ,
\end{eqnarray}
with lattice spacing $a$ and unit vector $\widehat{\vm}$ in the
$m$--direction. The change of the effective action is then
\begin{eqnarray} \label{DeltaGammageneral}
\Delta \Gamma\big[U\big] \equiv
\Gamma\big[U^\theta\big] - \Gamma\big[U\big]
                 =- \sum_n \ln \det \Big( \sum_{l\, }
        \mathcal{Q}^{(n)}_{kl}[U]\,\bar{\mathcal{Q}}^{(n)}_{lm}\Big),\,
\end{eqnarray}
with unitary transformation matrices
$\mathcal{Q}^{(n)}$ and $\bar{\mathcal{Q}}^{(n)}$.
For the case of odd $N$,
a long calculation gives for all $n \neq 0$:
\beq \ln \det
     \Big( \sum_{l\, }
     \mathcal{Q}^{(n)}_{kl}[U]\,\bar{\mathcal{Q}}^{(n)}_{lm}
     \Big)\, \Big|_{n \neq 0} =0\; .
\eeq
There remains the $n=0$ contribution \cite{KlinkhamerSchimmel}:
\beq\label{DeltaGamma}
 \Delta \Gamma[U] =  - \ln \det \Big(-
  a^3 \sum_{\vx} \psi^\dagger_k(\vx)\, W^{(0)}[U]\, \psi_m(\vx)\Big),
\eeq
with two-spinors $\psi_m(\mathbf{x})$ from an orthonormal basis and
a three-dimensional unitary operator
$ W^{(0)}$ (so that $\Delta \Gamma[U]$ is imaginary).

The determinant on the \rhs~of Eq.~(\ref{DeltaGamma}) is, in general,
unequal to 1 and the \DS{CPT} anomaly is seen to reduce effectively to
the three-dimensional  ``parity'' anomaly \cite{R84,ADM85,CL89},
as suggested by the heuristic argument of Sec.~\ref{Example}.
This establishes the four-dimensional \DS{CPT} anomaly for arbitrary $a$
and odd $N$. For  the case of even $N$, there is an additional determinant
(from the $n=N/2$ Fourier mode), which goes to 1 as $a \to 0$.

For $N=2$, it is, in fact, possible to calculate the
imaginary part of the effective action, not just the change under \DS{CPT}.
In the classical continuum limit $a \to 0$ (with
smooth $x_4$--independent gauge field $a_m(\vx)$ and $L'=N' a$ held fixed)
and with different charges $q_f$ present, the result is
\cite{KlinkhamerSchimmel}:
\begin{eqnarray}
\mathrm{Im}\,\Gamma^{(N=2)} [a\,] \sim
\Big({\textstyle \sum_f} \; q_f^2  \Big)
\big(2\pi + 0\big)\; \Omega_{\rm  CS}[a\,]\, ,\label{effactN2}
\end{eqnarray}
in terms of the \CS~integral
\begin{eqnarray}
\Omega_{\rm CS}[a\,]  \equiv \frac{1}{16\, \pi^2}
\int \mathrm{d}^3 x \; \epsilon^{klm}\, \partial_k a_l(\vx)\, a_m(\vx)\,.
\end{eqnarray}
The contribution $2\pi$ in the second factor on the \rhs~of
Eq.~(\ref{effactN2}) traces back to the $n=0$ Fourier
modes of the fermions and the contribution $0$ to the $n=1$ modes.

It would be of interest to calculate, either numerically or analytically,
$\mathrm{Im}\,\Gamma[a\,]$ for other simple setups,
preferably also with $x_4$--dependent gauge fields.

\section{Microcausality of the Maxwell--Chern--Simons model}
\label{Microcausality in 4D MCS}

For the four-dimensional Maxwell--\CS~(MCS) model (\ref{S-MCS})
in  the Coulomb gauge $\nabla \cdot \va=0$,
the following commutators of the electric field
$\ve \equiv \partial_0 \,\va -\nabla\, a_0$ and magnetic field
$\vb \equiv \nabla \times \va$ have been found \cite{AK01npb}:
\beqa
\hspace*{-7mm}
[e_i (x),e_j (0)]&\!=\!&
\Big( (\delta_{ij}\,\partial_0^2 -\partial_i \partial_j )\,(\partial_0^2 -\nabla^2 )
+ \msmall^2 \,\delta^3_i \delta^3_j \, \partial_0^2
 \no \\[0.1cm]
&&
- \,\msmall\, \epsilon_{ij3}\, \partial_0^3  + \msmall\,
( \epsilon_{ik3} \,\partial_j  - \epsilon_{jk3} \,\partial_i  )\,\partial_k \partial_0
\Big)  \, \ii\, D_\text{MCS}(x) ,\label{eq:comm-ee}
\\[0.2cm]
%%\eeqa\beqa
\hspace*{-7mm}
[e_i (x),b_j (0)] &\!=\!&
\Big( \epsilon_{ijk}\, \partial_k\partial_0 \,(\partial_0^2 -\nabla^2 )
 - \msmall^2 \, \delta^3_i\, \epsilon_{j3k}\, \partial_k \partial_0
\no \\[0.1cm]
&&
+ \,\msmall  \, \left( \delta_{ij}\,\partial_0^2  - \partial_i \partial_j\right)
\partial_3
-\msmall \,  \delta^3_j \,\partial_i (\partial_0^2 - \nabla^2 )
\Big) \,\ii\,  D_\text{MCS}(x) ,\label{eq:comm-eb}
\\[0.2cm]
%%\eeqa\beqa
[b_i (x), b_j (0)] &\!=\!&
\Big( (\delta_{ij}\nabla^2 -\partial_i \partial_j )\,(\partial_0^2 -\nabla^2 )
-\msmall \,\epsilon_{ijk}\, \partial_k \partial_0 \partial_3
\no \\[0.1cm]
&&
+\,\msmall^2 \,\Bigl( \delta_{ij}\,(\nabla^2 - \partial_3^2 )
- \partial_i\partial_j  - \delta^3_i  \delta^3_j\, \nabla^2
+( \delta^3_i\, \partial_j  + \delta^3_j\,\partial_i  )\, \partial_3 \Bigr)
\Big) \no\\
&&
\times \, \ii\,  D_\text{MCS}(x) , \label{eq:comm-bb}
\eeqa
with vector indices $i$, $j$, $k$ running over $1$, $2$, $3$,
natural units $\hbar$ $=$ $c$ $=$ $1$, and commutator function
\beq \label{eq:comm-func-space}
\hspace*{-1em}
D_\text{MCS}(x)
\equiv
(2\pi)^{-4}\,\oint_C \rmd p_0 \int \rmd^3 p\;
        \fracnew{  \exp \left[\,\ii\, p_0\, x^0 +\ii\, \vp \cdot \vx \,\right]  }
             {\left( p_0^2 - |\vp\,|^2 \right)^2 -
                \msmall^2 \left(p_0^2 - p_1^2 - p_2^2 \right)} \;\;,
\eeq
for a contour $C$ which encircles all four poles of the integrand
in the counterclockwise direction. Note that the derivatives on the
right-hand sides of Eqs.~(\ref{eq:comm-ee})--(\ref{eq:comm-bb})
effectively bring down powers of the momenta in the integrand of
Eq.~(\ref{eq:comm-func-space}).

The calculation of the commutators (\ref{eq:comm-ee})--(\ref{eq:comm-bb})
is rather subtle: $a_0$, for example, does not vanish in the Coulomb gauge
but is determined by a nondynamical equation,
$a_0 =\ii \,\msmall \,|\vp\,|^{-2}\,\epsilon_{3kl}\,a_k \,p_l$ in momentum space.
The Lorentz noninvariance of the MCS model is illustrated by the denominator of the
integrand in (\ref{eq:comm-func-space}) and the fact that, for example, the
commutators (\ref{eq:comm-ee}) and (\ref{eq:comm-bb}) differ at order $\msmall^2$.

Two further observations can be made.
First, the commutator function \eqref{eq:comm-func-space}
vanishes for spacelike separations,
\begin{equation}\label{spacelikecommutatorfunction}
D_\text{MCS}(x^0 ,\vx\,)=0\,, \quad \text{for}\;\; |x^0 |<|\vx\,| \; ,
\end{equation}
as follows by direct calculation.
Second, even though the commutators of the vector potentials $ \va\, (x)$
have  $|\vp\,|^{-2}$ poles which could potentially spoil causality, these poles
are absent for the commutators (\ref{eq:comm-ee})--(\ref{eq:comm-bb})
of the physical (gauge-invariant) electric and magnetic fields.
See Ref.~\cite{AK01npb} for further details.

The results \eqref{eq:comm-ee}--\eqref{spacelikecommutatorfunction}
establish microcausality of the MCS model (\ref{S-MCS}).
Apparently,
the well-known Jordan--Pauli field commutation relations of standard QED
\cite{JP28} (see also Refs.~\cite{Heitler1954,Gottfried1966})
can be deformed, at least in the way corresponding to the MCS model
with ``space\-like'' term  (\ref{CSterm}).
The space\-like MCS model with nonzero deformation parameter $m$
has, however, qualitatively different uncertainty relations
(\eg, a nonvanishing commutator of $b_1$ and $b_2$ fields
averaged over the same spacetime region).

The ``timelike'' MCS model, with $ \epsilon^{3\kappa\lambda\mu}$
in (\ref{CSterm}) replaced by $ \epsilon^{0\kappa\lambda\mu}$, does
violate microcausality, as long as unitarity is enforced \cite{AK01npb}.
This particular result may have other implications.
It rules out, for example, the possibility that a \CS-like
term can be radiatively induced from a \DS{CPT}--violating axial-vector
term in the Dirac sector \cite{AK01plb}.

\end{document}